\documentclass[usenatbib,fleqn, onecolumn]{mnras}

\usepackage{amsmath} 
\usepackage{amssymb}	
\usepackage{hyperref}
\usepackage{natbib}

\usepackage{graphicx}

\title{Temperature-dependent oscillation modes in rotating superfluid neutron stars}

\author[V.~A.~Dommes, E.~M.~Kantor, M.~E.~Gusakov]
{V.~A.~Dommes, E.~M.~Kantor, M.~E.~Gusakov
 \\
Ioffe Institute,
Polytekhnicheskaya 26, 194021 St.~Petersburg, Russia
}

\date{\today}

\renewcommand{\v}[1]{\ensuremath{\mbox{\boldmath$ #1 $}}}  
\newcommand{\abs}[1]{\left| #1 \right|} 
\renewcommand{\d}[2]{\frac{d #1}{d #2}} 
\newcommand{\pd}[2]{\frac{\upartial #1}{\upartial #2}} 
 
\begin{document}

\maketitle


\begin{abstract}
	We calculate the spectrum of inertial oscillation modes in
	a slowly rotating superfluid neutron star,
	including, for the first time, both the effects of finite temperatures
	and entrainment between superfluid neutrons and protons.
	We work in the Newtonian limit and assume minimal core composition
	(neutrons, protons and electrons).
	We also developed an approximate method
	that allows one to calculate the superfluid r-mode analytically.
	Finally, we derive and analyze dispersion relations for inertial modes 
	in the superfluid NS matter in the short wavelength limit.
	
\end{abstract}

\section{Introduction}
\label{sec:intro}

According to the standard r-mode theory, hot and rapidly rotating 
neutron stars (NSs) in low-mass X-ray binaries (LMXBs)
should be CFS-unstable with respect to emission of gravitational waves \citep{andersson98,fm98}. 
As a consequence, the probability of observing them should be very small,
but this conclusion contradicts observations \citep{hah11}.
A number of possible ideas have been proposed to solve the paradox
\citep[see, e.g.][]{ak01,hdh12, hah11,ms13},
but its complete resolution is still lacking.
\cite{gck14a,gck14b} 
introduced a new scenario, in which 
the finite-temperature effects in the superfluid core of an NS 
lead to resonance coupling and enhanced damping (and hence stability) 
of oscillation modes at certain ``resonance'' stellar temperatures.
It was demonstrated that NSs in LMXBs with high spin frequency may spend a substantial amount of time at these resonance temperatures, 
so that their interpretation does not constitute a problem.

The proposed resonance stabilization scenario 
was based on a simplified phenomenological model 
(in particular, resonance temperatures have never been explicitly calculated).
To put it on a more solid ground, one has to calculate
spectra of rotational inertial modes (modes for which the restoring force is the Coriolis force)
for realistic superfluid NS models
at arbitrary temperatures 
in order to find resonance temperatures at which the normal $r$-mode
exhibits an avoided crossing with another mode.
However, in most works \citep[e.g.][]{lm00,pca02,ly03,agh09} the inertial modes are
studied only in zero-temperature limit, when all neutrons and protons are assumed to be in a superfluid state.

\cite{kg17} considered normal and superfluid r-modes,
incorporating
finite-temperature effects and stratification by muons ($\mu$),
and
subsequently
found avoided crossings between normal and superfluid $r$-modes in the next-to-leading order in stellar rotation frequency.
This work ignored the entrainment between superfluid neutrons and protons, which significantly affects
the spectrum of superfluid inertial modes
\citep[see, e.g.][]{ly03}.
Also \cite{kg17} focused only on $r$-modes and have not studied other inertial modes, which also can interact with the normal $r$-mode.
To fill this gap,
we calculate the spectrum of inertial modes
in superfluid NSs whose cores consist of neutrons ($n$), protons ($p$) and electrons ($e$),
accounting for both entrainment and finite-temperature effects.
In Section~\ref{sec:eqns} we provide the equations governing these oscillation modes,
and in Section~\ref{sec:class} we discuss the general classification of inertial modes.
Section~\ref{sec:numerical} presents the results of numerical calculations for inertial modes,
obtained under assumptions of $npe$ NS core composition and constant critical temperatures.
In Section~\ref{sec:r-mode} we also present an approximate method that allows for calculation of the superfluid $r$-mode
analytically, in the limit of small entrainment.
In Section~\ref{sec:wkb} we derive dispersion relations for inertial modes in superfluid $npe$ matter
in short-wavelength limit, and explain some properties of these modes
that can be observed in the numerical results of Section~\ref{sec:numerical}.
Finally, we provide a summary in Section~\ref{sec:summary}.

\section{Equations governing oscillations of a rotating superfluid NS}
\label{sec:eqns}

In this section we describe oscillations of superfluid NS
using the Newtonian limit of relativistic hydrodynamics,  
formulated by \cite{gusakov16,gd16}.
We consider a slowly rotating (with the spin frequency $\Omega$) NS
with the core composed of neutrons, protons and electrons.
Possible superfluidity of baryons (neutrons and protons)
is encoded in the symmetric entrainment matrix $Y_{ik}$
\citep{ga06, gkh09a,gkh09b,ghk14}
which is the relativistic analogue of the superfluid mass-density matrix $\rho_{ik}$
\citep{ab76}.
This matrix enters the neutron ($i=\rm n$) and proton ($i=\rm p$) four-current density $j_{(i)}^\mu$:
\begin{gather}
	j_{(i)}^\mu  = n_i u^\mu + Y_{ik} w_{(k)}^\mu
.
\end{gather}
Here and hereafter indices $i,k$ run over neutrons and protons ($i,k = \rm n,p$),
and we assume summation over repeated indices.
$u^\mu$ is the four-velocity of the non-superfluid component
(electrons as well as baryonic thermal excitations),
and the four-vector $w_{(k)}^\mu$ is expressed through the superfluid velocity $v_{(sk)}^\mu$,
relativistic chemical potential $\mu_k$ and bare mass $m_k$ of particle species $k$
as $w_{(k)}^\mu = m_k v_{(sk)}^\mu - \mu_k u^\mu$.
Electron four-current is $j_{({\rm e})}^\mu = n_{\rm e} u^\mu$.
From the condition $j_{({\rm e})}^\mu = j_{({\rm p})}^\mu$,
which is valid for low-frequency hydrodynamic oscillations \citep{mendell91},
it follows
that $Y_{\rm pk} w_{(k)}^\mu = 0$.
Generally, $Y_{ik}$ depends on particle number densities $n_i$
and ratios $T/T_{{\rm c}i}$, where $T$ is the temperature and $T_{{\rm c}i}$
is the critical temperature for transition of particle species $i$ into the superfluid state.
The entrainment effect is described by off-diagonal entrainment matrix elements
$Y_{\rm np} = Y_{\rm pn}$.

Below we write down the linearized equations describing small oscillations,
with all perturbations depending on time as ${\rm e}^{\imath \sigma t}$
in the frame rotating with the star (with frequency $\Omega$).
In the Newtonian limit and assuming Cowling approximation these equations consist of \citep{kg17}:

(i) Continuity equations for baryons and electrons
\begin{gather}
\label{eq:general-cont}
\delta n_{\rm b}+{\rm div}(n_{\rm b} \v{\xi}_{\rm b}) = 0,
\quad
\delta n_{\rm e}+{\rm div}(n_{\rm e} \v{\xi}) = 0.
\end{gather}
Here and hereafter $\delta$ stands for the Euler perturbation
of the corresponding thermodynamic quantity;
$\v{\xi} \equiv \v{j}_{\rm e}/(\imath \sigma n_{\rm e})$
is the Lagrangian displacement of the normal liquid component
(electrons as well as non-superfluid neutrons and protons)
\footnote{We assume that all non-superfluid components move
with one and the same velocity due to efficient particle collisions.},
and 
$\v{\xi}_{\rm b} \equiv \v{j}_{\rm b}/(\imath \sigma n_{\rm b})$ 
is the Lagrangian displacement for baryons;
$n_{\rm b}\equiv n_{\rm n}+n_{\rm p}$ and $\v{j}_{\rm b} \equiv \v{j}_{\rm n} + \v{j}_p$
are the baryon number density and baryon current density, respectively.

(ii) Euler equation
\begin{gather}
\label{eq:general-euler}
	-\sigma^2 { \v{\xi}_{\rm b}}+2 \imath \sigma \v{\Omega}\times {\v{\xi}_{\rm b}}
	 = \frac{\delta w}{w^2} \v{\nabla} P-\frac{\v{\nabla} \delta P}{w}
,
\end{gather}
where $w=(P+\epsilon)/c^2$, 
$P$ is the pressure, $\epsilon$ is the energy density, 
and $c$ is the speed of light;

(iii) the `superfluid' equation, analogue of the Euler equation 
for superfluid (neutron) liquid component,
\begin{gather}
\label{eq:general-sfl}
	h \sigma^2 \v{z}- 2\imath h_1 \sigma \v{\Omega} \times \v{z}
	= c^2 n_{\rm e} \v{\nabla} \Delta \mu_{\rm e}
.
\end{gather}
Here $\v{z}\equiv \v{\xi}_{\rm b} - \v{\xi}$ is the superfluid Lagrangian displacement,
$\Delta \mu_{\rm e} \equiv \mu_{\rm n}-\mu_{\rm p}-\mu_{\rm e}$
is the chemical potential imbalance (note that $\delta \Delta \mu_{\rm e}=\Delta \mu_{\rm e}$
since for an unperturbed star $\Delta \mu_{\rm e}=0$ due to the condition of beta-equilibrium),
and quantities $h$ and $h_1$ are expressed through the entrainment matrix $Y_{ik}$ as
\begin{gather}
\label{eq:h}
	h=n_{\rm b} \mu_{\rm n} y
,\quad
	h_1=\mu_{\rm n} n_{\rm b}\left(\frac{n_{\rm b}}{Y_{{\rm nn}}\mu_{\rm n} + Y_{{\rm np}}\mu_{\rm p}}-1\right)
,\quad
	y=\frac{n_{\rm b}Y_{\rm pp}}{\mu_{\rm n}(Y_{\rm nn}Y_{\rm pp}-Y_{\rm np}^2)}-1
.
\end{gather}
Note that in the absence of entrainment ($Y_{\rm np} = 0$) $h_1$ and $h$ are equal, $h_1 = h$.
The `superfluid' equation takes the form \eqref{eq:general-sfl}
if the interaction between the neutron vortices and normal component is weak (the weak-drag regime),
which is true for typical NS conditions 
\citep[see, e.g.,][]{mendell91, asc06}.

The equations (i)-(iii) should be supplemented by the `equation of state' (EOS),
$\delta n_i
=\frac{\partial n_i}{\partial P} \delta P
+\frac{\partial n_i}{\partial \Delta \mu_{\rm e}} \Delta \mu_{\rm e}
.$

In the present study we are interested in 
the inertial oscillation modes, 
whose eigenfrequencies $\sigma$ 
vanish as $\Omega \rightarrow 0$. 
Thus, up to the terms $\sim (\Omega/\Omega_0)^2$ ($\Omega_0$ is of the order of Kepler frequency), 
the eigenfrequency 
$\sigma$, the Euler perturbation 
of any (scalar) thermodynamic parameter $f$ 
(e.g., $P$, $\mu_{\rm e}$, $n_{\rm b}$ etc.),
and the Lagrangian displacement $\v{d}$ (e.g., $\v{\xi}_{\rm b}$ or $\v{z}$)
can be presented as \citep[e.g.,][]{pbr81,lf99}
\begin{gather}
\label{eq:sigma}
\sigma=\Omega \sigma_0\left(1+\Omega^2 \sigma_1\right)
,\\
\delta f(t,r,\theta,\phi) =\Omega^2 \delta f^1(r,\theta) {\rm exp}(\imath \sigma t+\imath m \phi)
\label{eq:deltaf}
,\\
\label{eq:d}
\v{d}(t,r,\theta,\phi) = \left[\v{d}^0(r,\theta) + \Omega^2 \v{d}^1(r,\theta) \right] {\rm exp}(\imath \sigma t+\imath m \phi)
,
\end{gather}
where $m$ is an integer and $(r,\theta,\phi)$ are spherical coordinates
with the origin at the stellar center.

In this paper we work in the leading order in $\Omega/\Omega_0$, i.e.
ignore
the terms $\sigma_1$, $\v{d}^1$ and the stellar oblateness
(i.e. all the equilibrium quantities depend on the radial coordinate only).
Then the system (i)-(iii) can be represented, in spherical coordinates, as
\begin{gather}
\label{eq:cont-b}
	\frac{1}{n_{\rm b}} \frac{1}{r^2}\pd{}{r}r^2 n_{\rm b} \xi_{{\rm b} r}^0
	+ \frac{1}{r\sin\theta}
	\left(
		\pd{}{\theta} \sin\theta \xi_{{\rm b} \theta}^0
		+ \imath m \xi_{{\rm b} \phi}^0
	\right)
	= 0
,\\
\label{eq:cont-e}
	\frac{1}{n_{\rm e}} \frac{1}{r^2}\pd{}{r}r^2 n_{\rm e} \left( \xi_{{\rm b}r}^0 - z_r^0 \right)
		+ \frac{1}{r\sin\theta}
	\left[
		\pd{}{\theta} \sin\theta \left( \xi_{{\rm b}\theta}^0 - z_\theta^0 \right)
		+ \imath m \left( \xi_{{\rm b} \phi}^0 - z_\phi^0 \right)
	\right]
	= 0
,\\
\label{eq:euler-r}
	- \sigma_0^2 \xi_{{\rm b}r}^0 - 2 \imath \sigma_0 \sin\theta \xi_{{\rm b}\phi}^0
	= - \pd{}{r} \frac{\delta P^1}{w}
	+ \frac{\mu_{\rm n}}{w^2 c^2}
	  		\pd{n_{\rm b}}{\Delta\mu_{\rm e}} \Delta\mu_{\rm e}^1
	  	\d{P}{r}
,\\
\label{eq:euler-th}
	- \sigma_0 \xi_{{\rm b}\theta}^0 - 2 \imath \cos\theta \xi_{{\rm b}\phi}^0
	= \frac{1}{\imath m} \pd{}{\theta} 
	\sin\theta
	\left[
		- \sigma_0 \xi_{{\rm b}\phi}^0 + 2 \imath \left( \xi_{{\rm b}r}^0 \sin\theta + \xi_{{\rm b}\theta}^0 \cos\theta \right) 
	\right]
,\\
\label{eq:euler-phi}
	- \sigma_0^2 \xi_{{\rm b}\phi}^0 + 2 \imath \sigma_0
		\left( \xi_{{\rm b}r}^0 \sin\theta + \xi_{{\rm b}\theta}^0 \cos\theta \right) 
	= -  \frac{1}{w} \frac{\imath m }{r \sin\theta} \delta P^1
,\\
\label{eq:sfl-r}
	- \sigma_0^2 z_r^0 - 2 \imath \frac{h_1}{h} \sigma_0 \sin\theta z_\phi^0
	= - \frac{c^2 n_{\rm e}}{h} \pd{}{r} \Delta \mu_{\rm e}^1
,\\
\label{eq:sfl-th}
	- \sigma_0 z_\theta^0 - 2 \imath \frac{h_1}{h} \cos\theta z_\phi^0
	= \frac{1}{\imath m} \pd{}{\theta} 
	\sin\theta
	\left[
		- \sigma_0 z_\phi^0 + 2 \imath \frac{h_1}{h} \left( z_r^0 \sin\theta + z_\theta^0 \cos\theta \right)
	\right]
,\\
\label{eq:sfl-phi}
	- \sigma_0^2 z_\phi^0 + 2 \imath \frac{h_1}{h} \sigma_0 \left( z_r^0 \sin\theta + z_\theta^0 \cos\theta \right)
		 = - \frac{c^2 n_{\rm e}}{h} \frac{\imath m}{r \sin\theta} \Delta \mu_{\rm e}^1
.
\end{gather}
To obtain these equations we
(i) substituted \eqref{eq:sigma}--\eqref{eq:d} into equations \eqref{eq:general-cont}--\eqref{eq:general-sfl},
(ii) omitted higher-order in $\Omega/\Omega_0$ terms,
(iii) substituted $\delta P^1$ from the $\phi$-component of the Euler equation \eqref{eq:euler-phi} into
the $\theta$-component of the Euler equation,
(iv) substituted  $\Delta\mu_{{\rm e}}^1$ from the $\phi$-component of superfluid equation \eqref{eq:sfl-phi} into
the $\theta$-component of the superfluid equation,
(v) expressed $\delta w$ in \eqref{eq:euler-r} through $\delta P$ and $\Delta\mu_{{\rm e}}$ \citep[see][Appendix A]{kg17}, and
(vi) divided superfluid equation by $-h(r)$.

It is convenient to express the functions
$\xi_{{\rm b}\theta}^0$,
$\xi_{{\rm b}\phi}^0$,
$z_\theta^0$, and $z_\phi^0$
in the system~\eqref{eq:cont-b}--\eqref{eq:sfl-phi}
as a sum of toroidal ($T$, $T_z$) and poloidal ($Q$, $Q_z$) components
\citep{saio82}:
\begin{gather}
\label{eq:xib-tp}
	\xi_{{\rm b}\theta}^0 = \frac{\partial}{\partial \theta}Q(r,\theta)+\frac{\imath m T(r,\theta)}{{\rm sin}\theta}
,\quad
	\xi_{{\rm b}\phi}^0 = \frac{\imath m Q(r,\theta)}{{\rm sin}\theta} -\frac{\partial}{\partial \theta}T(r,\theta)
,\\
\label{eq:z-tp}
	z_\theta^0 = \frac{\partial}{\partial \theta}Q_z(r,\theta)+\frac{\imath m T_z(r,\theta)}{{\rm sin}\theta}
,\quad
	z_\phi^0 = \frac{\imath m Q_z(r,\theta)}{{\rm sin}\theta} -\frac{\partial}{\partial \theta}T_z(r,\theta).
\end{gather}
Then, following the same procedure as for non-superfluid stars \citep[e.g.,][]{lf99}, we expand all the unknown functions
into Legendre polynomials with fixed $m$:
\begin{gather}
\label{eq:lm-xib}
	\xi_{{\rm b} r}^0(r,\theta)=\sum_{l_2}\xi_{{\rm b} r\, l_2m}^0(r) P_{l_2}^m(\cos\theta)
,\\
\label{eq:lm-z}
	z_{r}(r,\theta)=\sum_{l_2}z_{r\, l_2m}^0(r) P_{l_2}^m(\cos\theta)
,\\
\label{eq:lm-Q}
	Q(r,\theta)=\sum_{l_2}Q_{l_2m}(r) P_{l_2}^m(\cos\theta)
,\\
\label{eq:lm-Qz}
	Q_z(r,\theta)=\sum_{l_2}Q_{z\,l_2m}(r) P_{l_2}^m(\cos\theta)
,\\ 
\label{eq:lm-T}
	T(r,\theta)=\sum_{l_1}T_{l_1m}(r) P_{l_1}^m(\cos\theta)
,\\
\label{eq:lm-Tz}
	T_z(r,\theta)=\sum_{l_1}T_{z\,l_1m}(r) P_{l_1}^m(\cos\theta)
,\\
\label{eq:lm-dP}
	\delta P^1(r,\theta)=\sum_{l_2}\delta P_{l_2m}^1(r) P_{l_2}^m(\cos\theta)
,\\
\label{eq:lm-dmu}
	\Delta \mu_{{\rm e}}^1(r,\theta)=\sum_{l_2}\Delta \mu_{{\rm e}\,l_2m}^1(r) P_{l_2}^m(\cos\theta), 
\end{gather}
where the summation goes over $l_1 = m+2k$ and $l_2 = m+2k+1$ ($k=0,1,2,\ldots$)
for `odd' modes, and over $l_1 = m+2k+1$, $l_2 = m+2k$ for `even' modes.\footnote{
Following \cite{yl00}, we call `even' the modes whose
scalar perturbations are symmetric with respect to the equator,
and `odd' -- the modes with asymmetric perturbations.
Odd and even modes are completely decoupled and do not mix with each other.
}
After substituting these expansions into oscillation equations,
one obtains an infinite set of ordinary differential equations for unknown functions 
$\xi_{{\rm b} r\, l_2m}^0 (r)$, $z_{r\, l_2m}^0(r)$, \ldots

The oscillation equations should be supplemented by the boundary conditions, which consist of:

(i) regularity condition for the perturbations in the stellar center,

(ii) vanishing of the Lagrangian perturbation of the pressure
at the stellar surface $r=R$,
\begin{gather}
	\delta P(R) + \xi_{{\rm b} r}(R) \left.\d{P}{r}\right|_{r=R} = 0,
\end{gather}

(iii) continuity of $\delta P$, $\xi_{{\rm b} r}$ and $\xi_r$
at the superfluid/non-superfluid interface.

\section{Classification of rotational modes}
\label{sec:class}

We consider rotational oscillation modes with $\sigma \propto \Omega$
in the slow-rotation approximation
\citep{lf99,yl00}.
Each mode is characterized by two angular `quantum numbers', $l_0$ and $m$,
where $m$ is azimuthal number and $l_0$ (in the notation by \citealt{li99, yl00})
is the maximum index $l$ of spherical harmonics associated with the dominant expansion coefficients
of the eigenfunctions.
For the uniform density stars all coefficients with $l > l_0$ are strictly zero \citep{lf99}.

For a given $m$, there are two nodeless modes with $l_0 - |m|  = 1$:
the purely toroidal normal r-mode with $\sigma_0=2/(m+1)$,
and the superfluid r-mode, which, in the limit $Y_{\rm np} = 0$,
is also purely toroidal and has the same frequency \citep{ac01,ly03,agh09,kg17}.
For a given $m$ and $l_0  > |m| + 1$
there are $l_0 - |m|$ normal inertial modes ($i^o$-modes)
and $l_0 - |m|$ superfluid inertial modes ($i^s$-modes).
The modes where normal and superfluid components are comoving,
so that $|\v{\xi}_{\rm b}| \sim |\v{z}|$, are called `normal', or `ordinary', and denoted with a superscript $^o$;
If normal and superfluid components are counter-moving, so that 
the total baryon current is almost not excited, $|\v{\xi}_{\rm b}| \ll |\v{z}|$,
then the corresponding modes are referred to as `superfluid', and designated with a superscript $^s$.
The number of radial nodes in eigenfunctions of a given mode is determined by $l_0$ and $m$ (see \citealt{yl00}, Table 3).
For example, the dominant lowest-order toroidal eigenfunction [$T_{mm}(r)$ for $i^o$-modes, and $T_{z\,mm}(r)$ for $i^s$-modes]
has no nodes for the $r$-mode ($l_0 - |m| = 1$), one node for $l_0 - |m| = 3$ mode, and two nodes for $l_0 - |m| = 5$ mode.

\begin{figure}
	\centering
	\begin{minipage}{.48\textwidth}
		\centering
		\includegraphics[width=1.\linewidth]{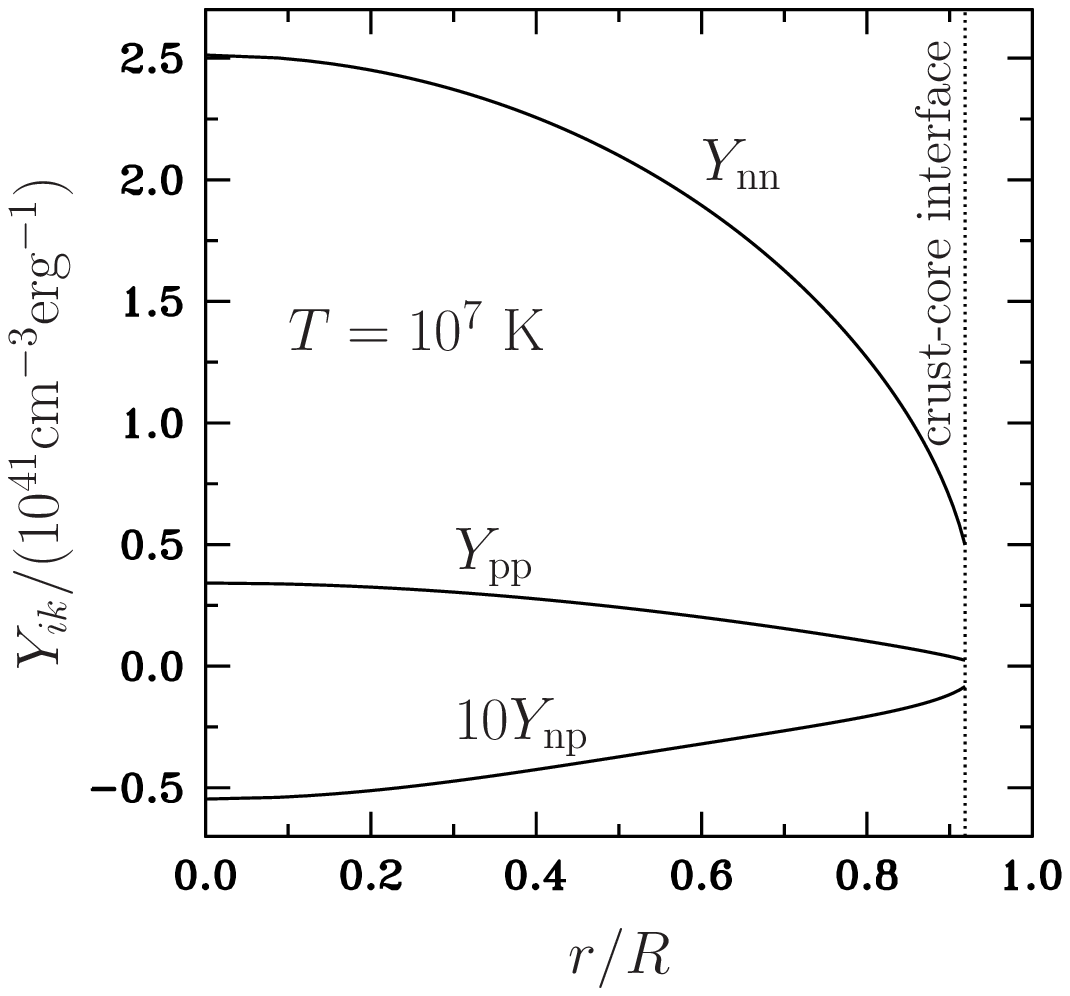}
		\label{fig:Yik:R}
	\end{minipage}
	\begin{minipage}{.48\textwidth}
		\centering
		\includegraphics[width=1.\linewidth]{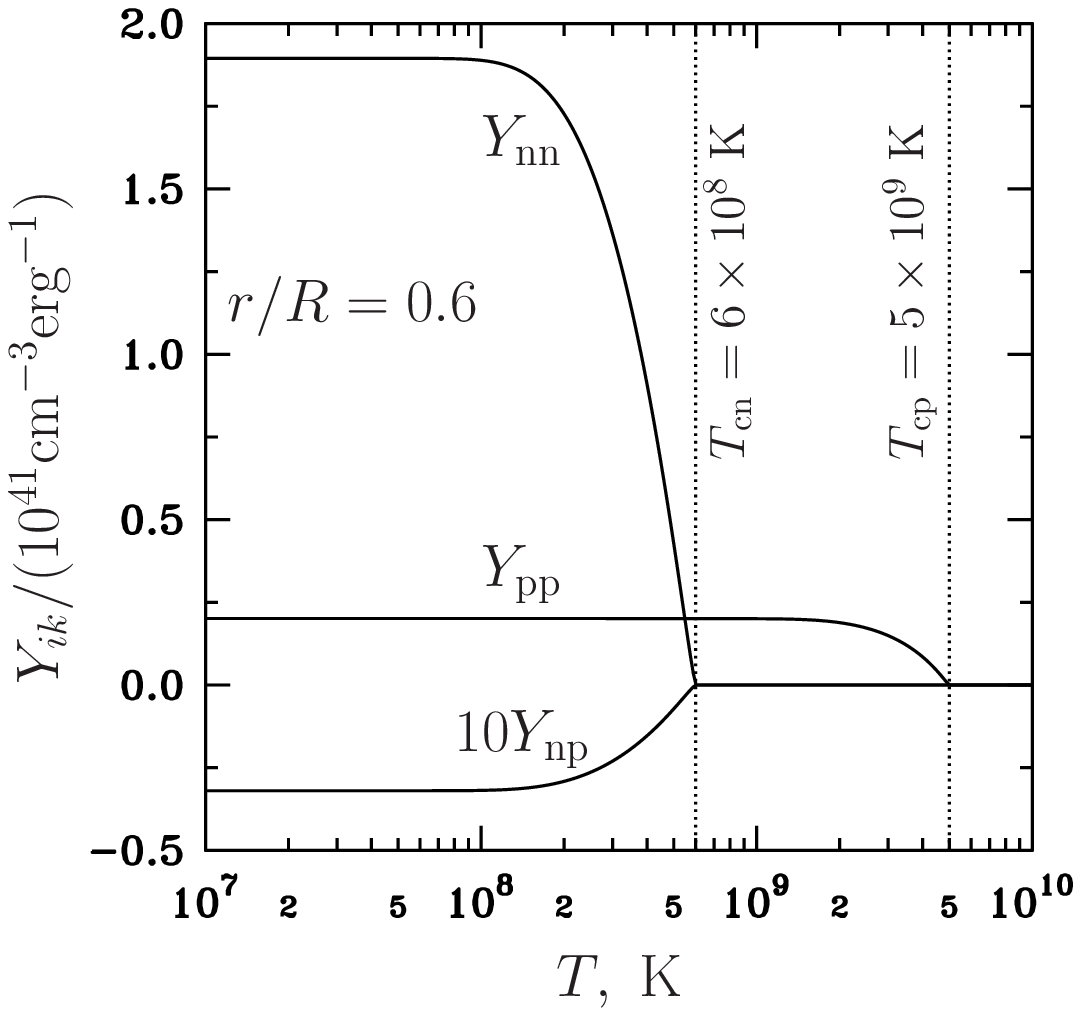}
		\label{fig:Yik:T}
	\end{minipage}
	
	\caption{
		Elements of the entrainment matrix $Y_{ik}$
		(in units of $10^{41}~{\rm cm}^{-3} {\rm erg}^{-1}$)
		versus $r/R$ at $T = 10^7~{\rm K}$ (left panel)
		and versus temperature $T$ at $r/R = 0.6$ (right panel).
		Critical temperatures are constant throughout the core,
		$T_{\rm cn} = 6 \times 10^8~{\rm K}$, $T_{\rm cp} = 5 \times 10^9~{\rm K}$.
	}
	\label{fig:Yik}
\end{figure}
\begin{figure}
\centering
    \includegraphics[scale=1.0]{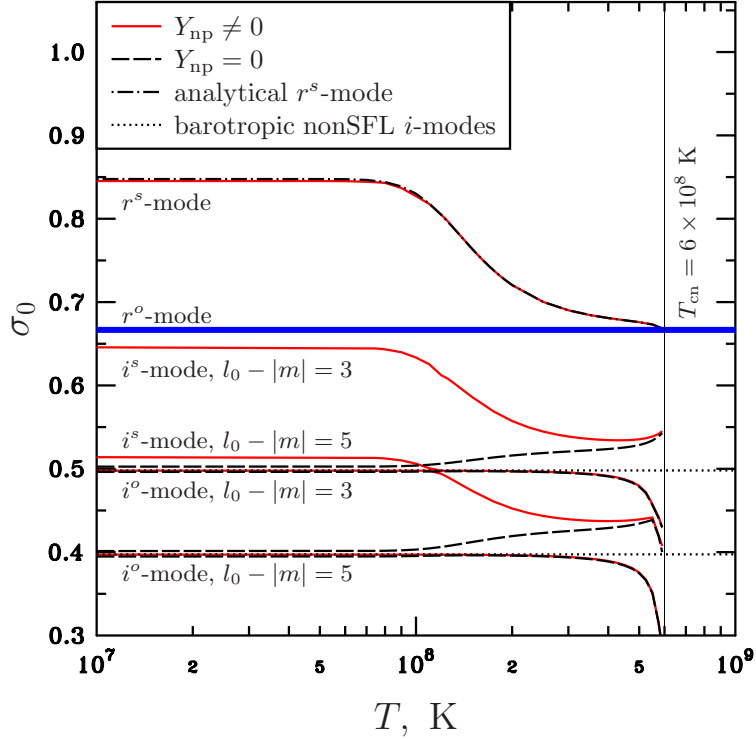}
    \caption{
    Eigenfrequency $\sigma_0$ versus stellar temperature $T$ for $m=2$ inertial modes.
    Critical temperatures are constant throughout the core,
    $T_{\rm cn} = 6 \times 10^8~{\rm K}$, $T_{\rm cp} = 5 \times 10^9~{\rm K}$.
    Solid lines denote inertial modes calculated taking into account the entrainment effect ($Y_{\rm np} \neq 0$),
    dashed lines denote the same modes calculated without the entrainment effect.
    Dot-dashed line denotes the superfluid $r^s$-mode calculated via the approximate analytical method
    introduced in Section~\ref{sec:r-mode}. 
    Dotted lines correspond to $l_0 - |m| = 3$ and $l_0 - |m| = 5$ normal $i$-modes
    in a non-superlfuid barotropic NS with the same EOS and mass.
    In all cases the $m=2$ normal $r$-mode (bold line) has the same frequency $\sigma_0 = 2/3$.
    } \label{fig:spectrumnpe}
\end{figure}

\begin{figure}
\centering
  \includegraphics[scale=1.0]{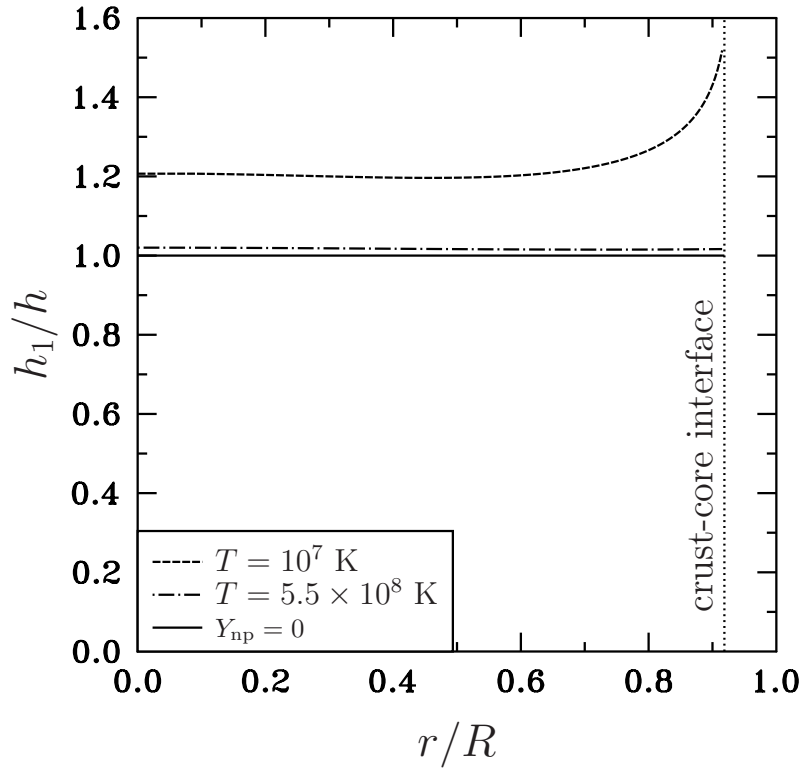}
\caption{
The ratio $h_1/h$ versus normalized radial coordinate $r/R$
for temperatures $T = 10^7~{\rm K}$ (dashed line)
and $T = 5.5 \times 10^8~{\rm K}$ (dot-dashed line).
Critical temperatures are constant throughout the core,
$T_{\rm cn} = 6 \times 10^8~{\rm K}$, $T_{\rm cp} = 5 \times 10^9~{\rm K}$.
In the absence of entrainment ($Y_{\rm np} = 0$) $h_1/h \equiv 1$ (solid line).
}
\label{fig:h1h}
\end{figure}

\begin{figure}
\centering
    \includegraphics[scale=1.0]{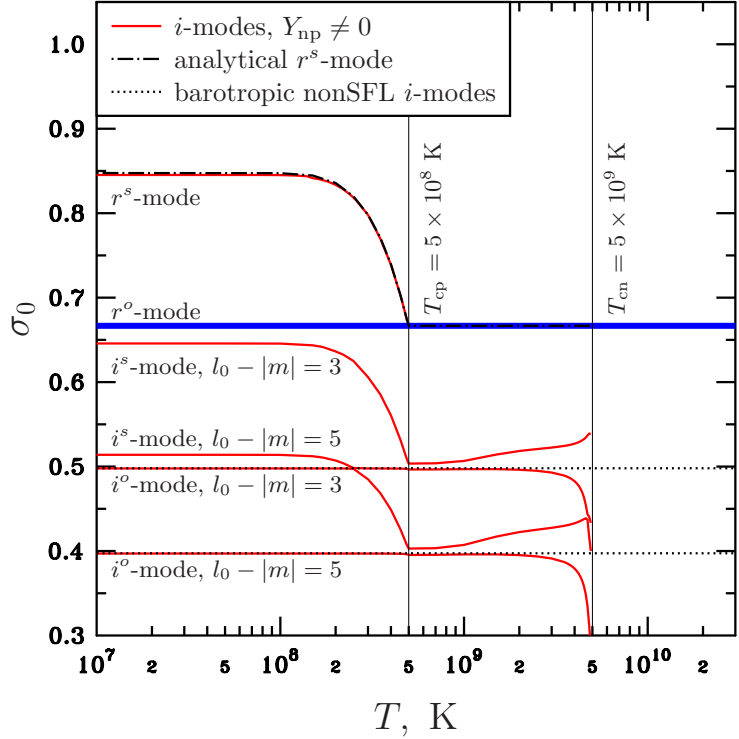}
    \caption{
    Eigenfrequency $\sigma_0$ versus stellar temperature $T$ for $m=2$ inertial modes.
    Critical temperatures are constant throughout the core,
    $T_{\rm cn}=5 \times 10^9\,\rm K$, $T_{\rm cp}=5 \times 10^8\,\rm K$.
    Solid lines denote inertial modes calculated taking into account the entrainment effect ($Y_{\rm np} \neq 0$ at $T < T_{\rm cp}$),
    Dot-dashed line denotes the superfluid $r^s$-mode calculated via the approximate analytical method
    introduced in Section~\ref{sec:r-mode}. 
    Dotted lines correspond to $l_0 - |m| = 3$ and $l_0 - |m| = 5$ normal $i$-modes
    in a non-superlfuid barotropic NS with the same EOS and mass.
    In all cases the $m=2$ normal $r$-mode (bold line) has the same frequency $\sigma_0 = 2/3$.
    } \label{fig:spectrumnpe-lowTcp}
\end{figure}

\begin{figure}
\centering
\begin{minipage}{.32\textwidth}
  \centering
  \includegraphics[width=1.\linewidth]{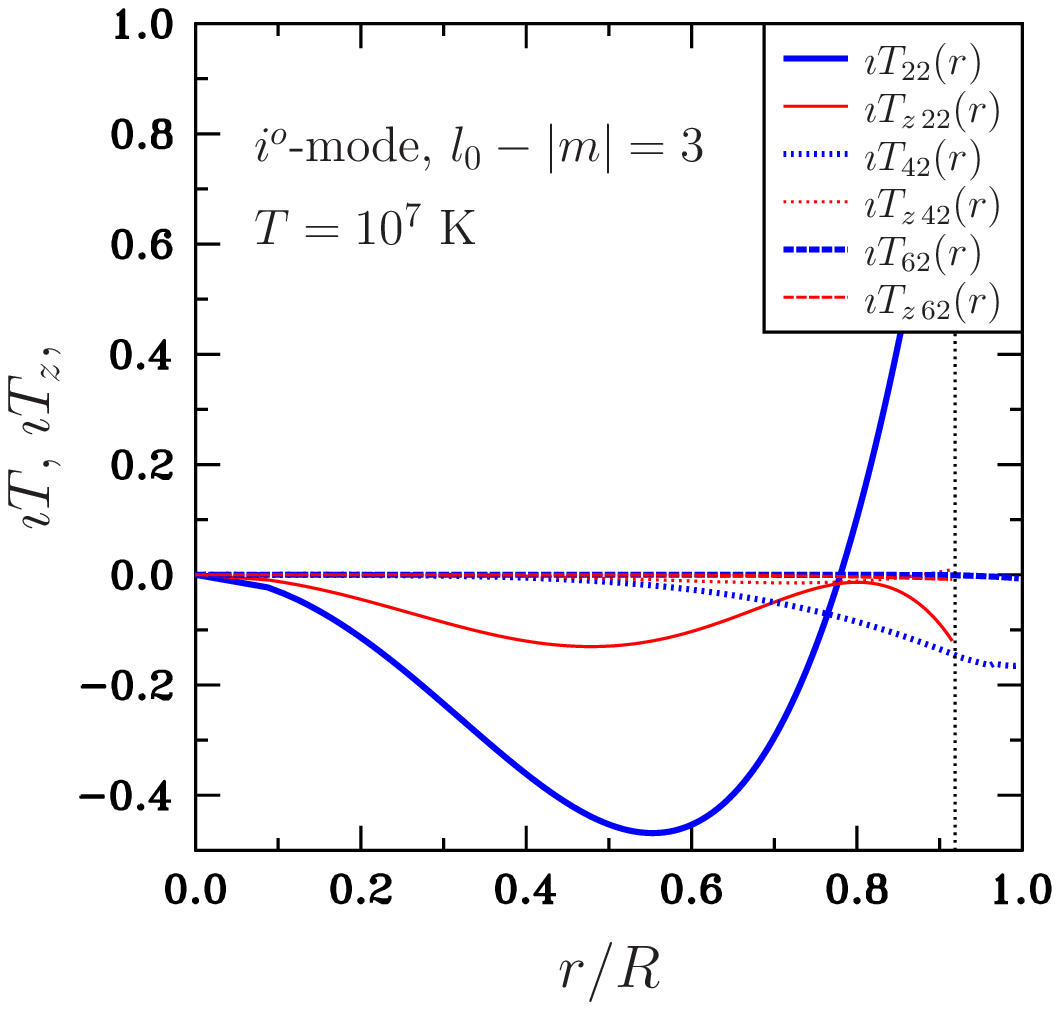}
\end{minipage}
\begin{minipage}{.32\textwidth}
  \centering
  \includegraphics[width=1.\linewidth]{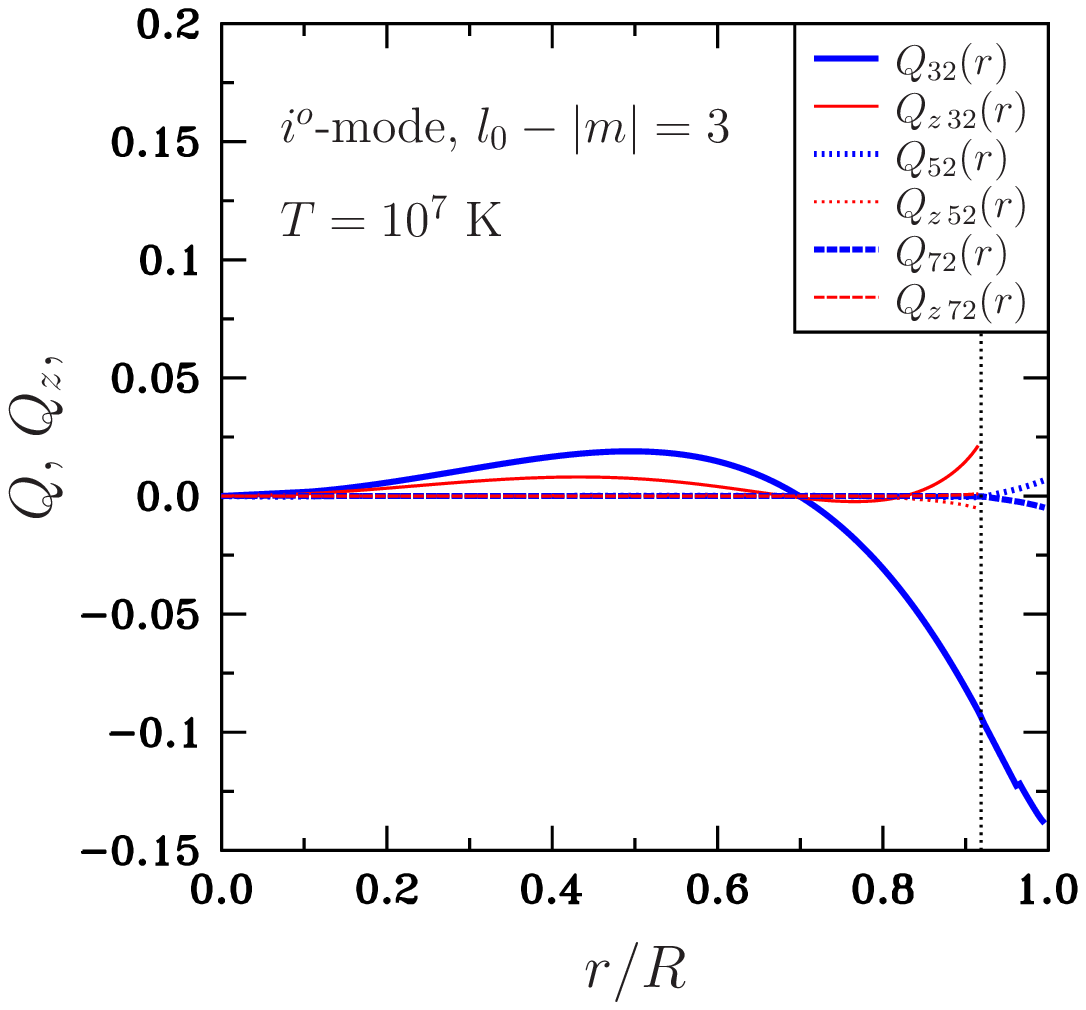}
\end{minipage}
\begin{minipage}{.32\textwidth}
  \centering
  \includegraphics[width=1.\linewidth]{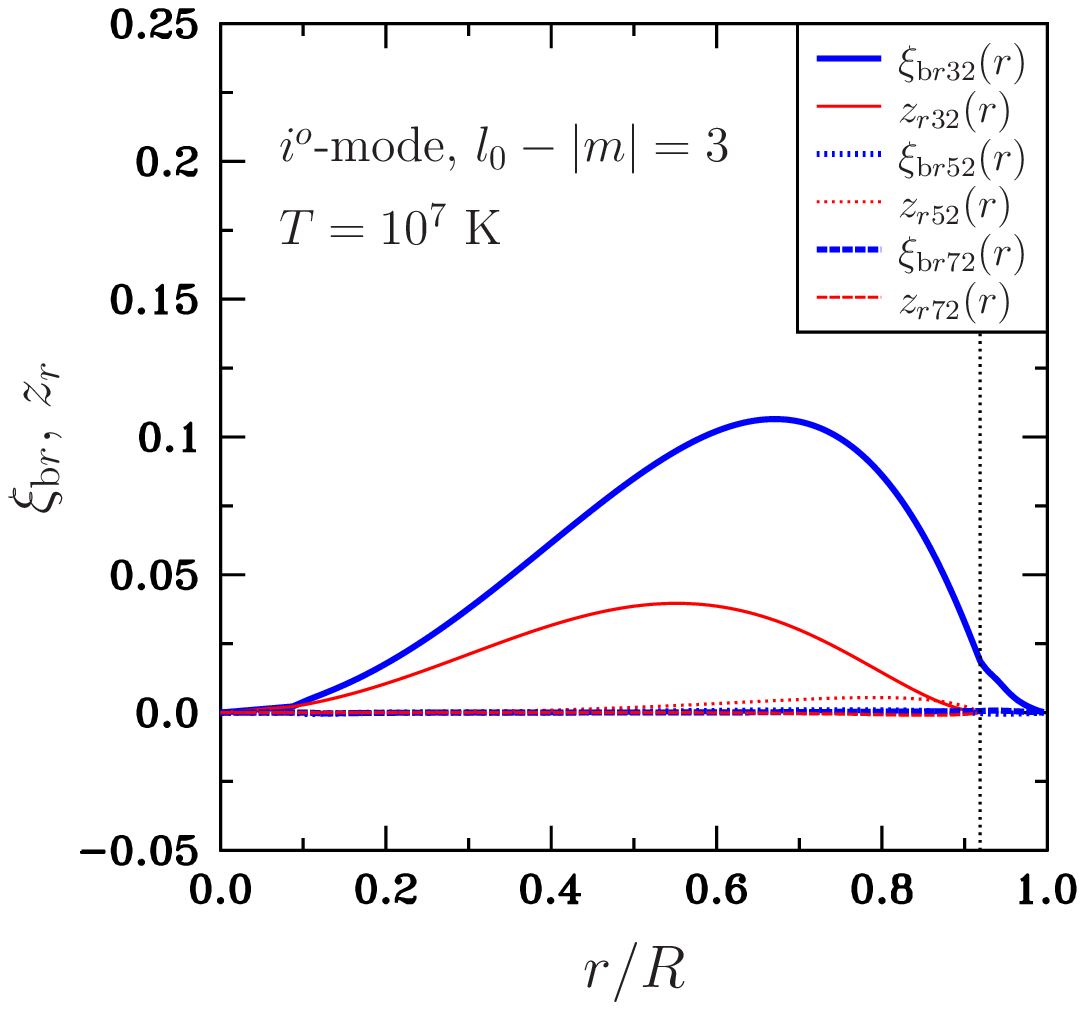}
\end{minipage}
\caption{
The lowest three harmonics
for toroidal (left panel), poloidal (central panel) and radial (right panel) displacements
for the $m=2$, $l_0 - \abs{m} = 3$ normal $i^o$-mode at $T=10^7~{\rm K}$.
Bold lines denote Lagrangian displacements for baryons, and thin lines denote superfluid displacements.
Critical temperatures are constant throughout the core,
$T_{\rm cn} = 6 \times 10^8~{\rm K}$, $T_{\rm cp} = 5 \times 10^9~{\rm K}$.
Vertical dots show the crust-core interface.
}
\label{fig:eigenfuncs:i3}
\end{figure}

\begin{figure}
\centering
\begin{minipage}{.48\textwidth}
  \centering
  \includegraphics[width=1.\linewidth]{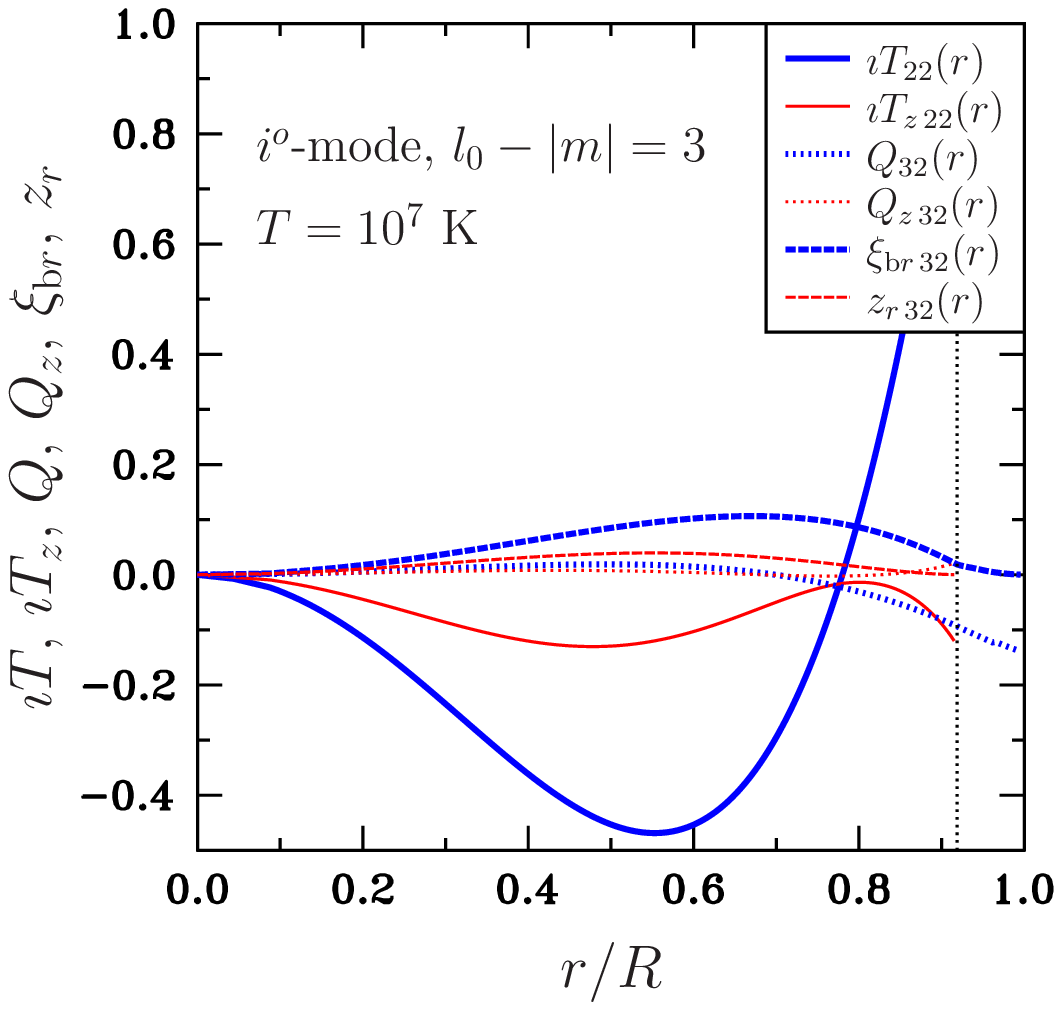}
  \label{fig:eigenfuncs:io3}
\end{minipage}
\begin{minipage}{.48\textwidth}
  \centering
  \includegraphics[width=1.\linewidth]{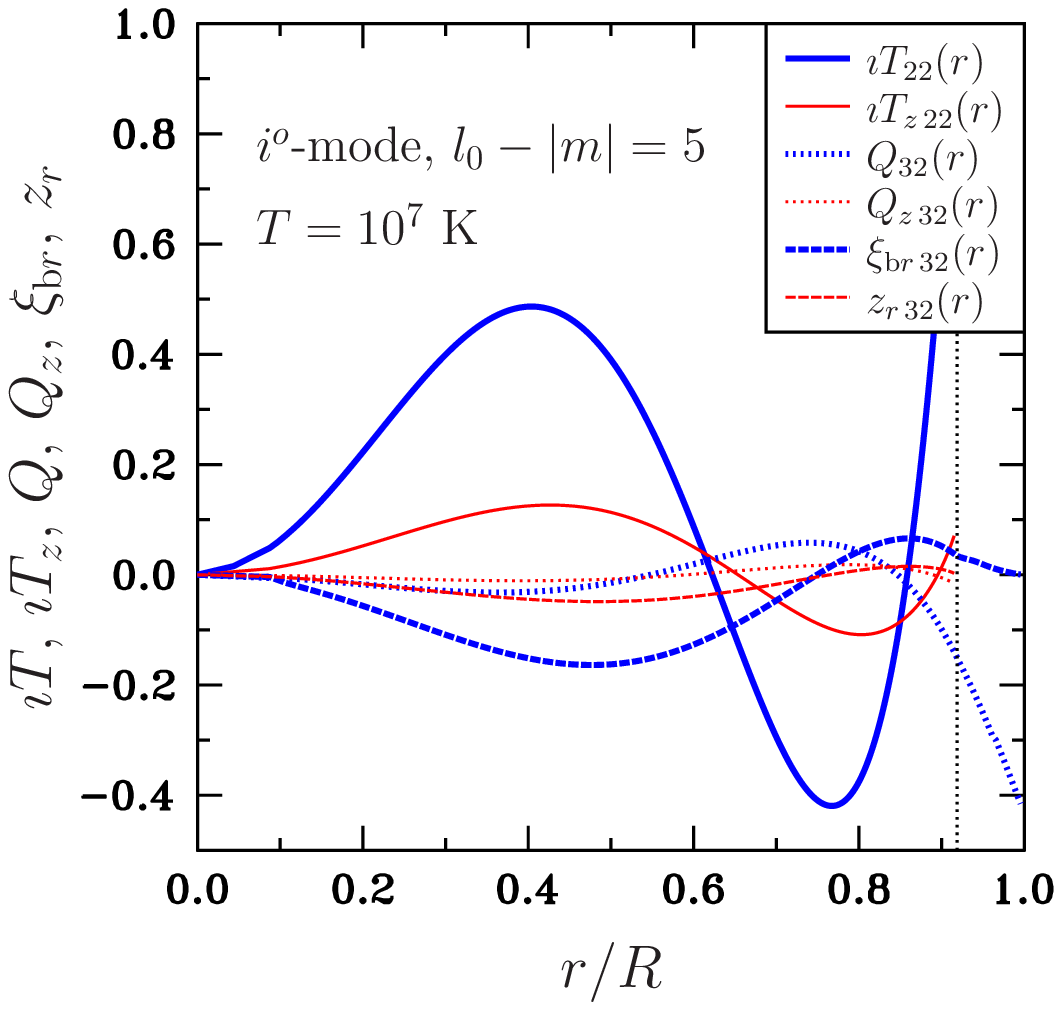}
  \label{fig:eigenfuncs:io5}
\end{minipage}
\begin{minipage}{.48\textwidth}
  \centering
  \includegraphics[width=1.\linewidth]{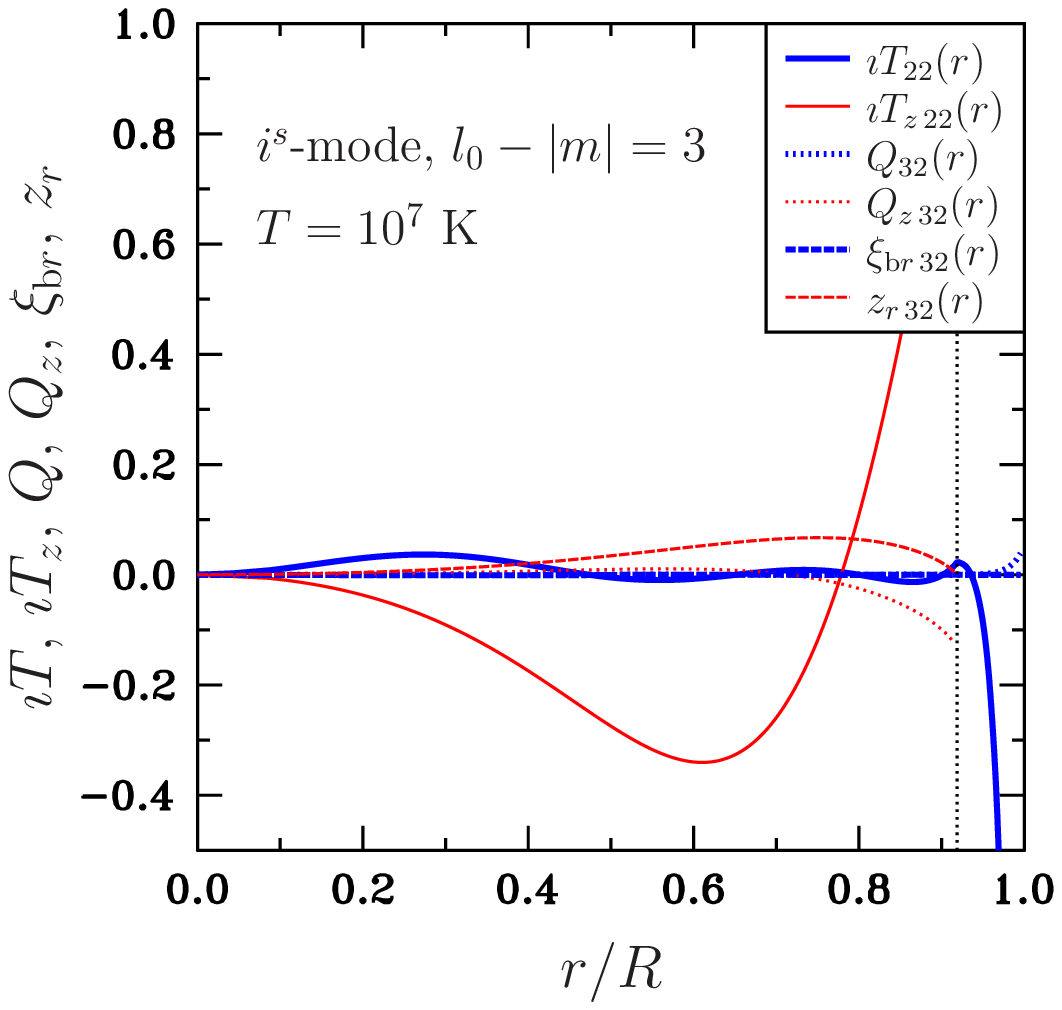}
  \label{fig:eigenfuncs:is3}
\end{minipage}
\begin{minipage}{.48\textwidth}
  \centering
  \includegraphics[width=1.\linewidth]{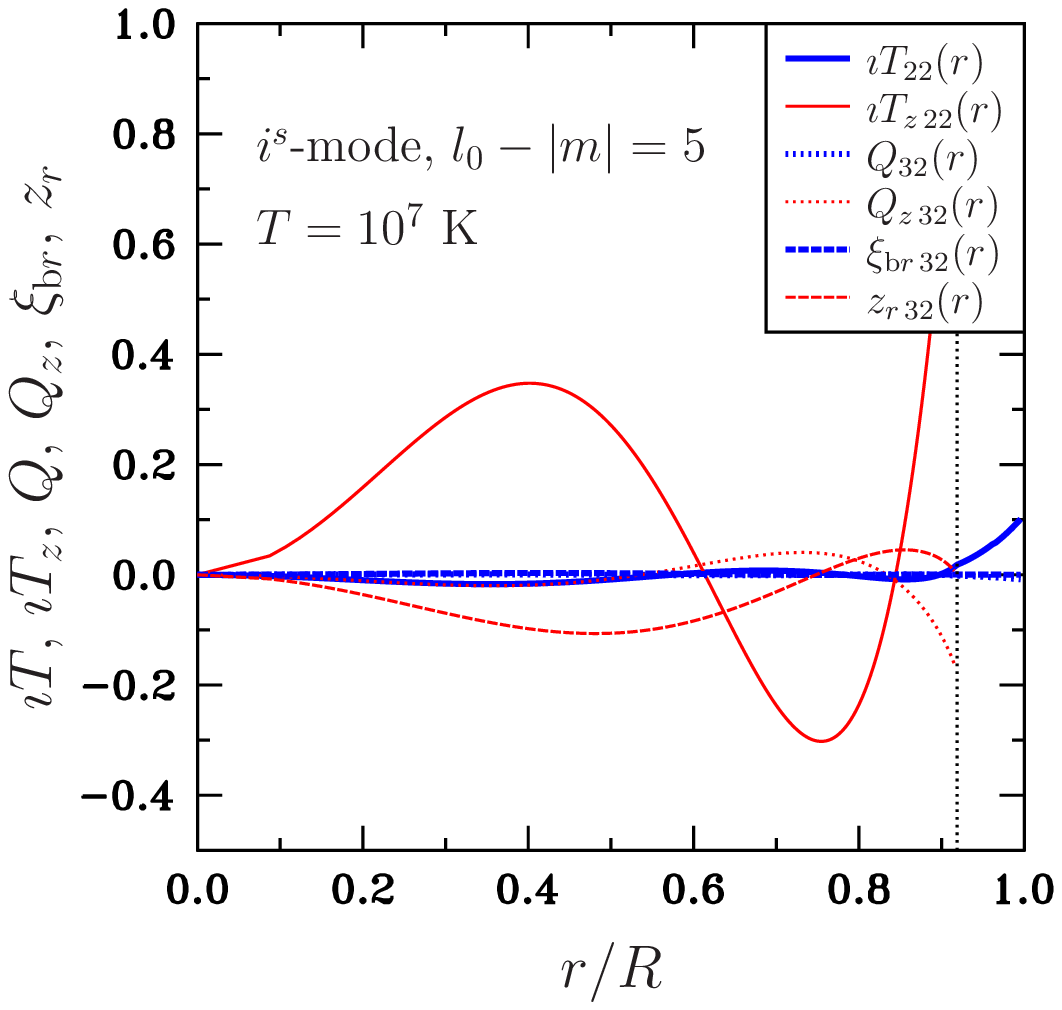}
  \label{fig:eigenfuncs:is5}
\end{minipage}

\caption{
Eigenfunctions for different $m=2$ inertial modes at $T = 10^7~{\rm K}$.
Top left: $l_0 - \abs{m} = 3$ normal $i^o$-mode.
Top right: $l_0 - \abs{m} = 5$ normal $i^o$-mode.
Bottom left: the $l_0 - \abs{m} = 3$ superfluid $i^s$-mode.
Bottom right: the $l_0 - \abs{m} = 5$ superfluid $i^s$-mode.
Only lowest-order harmonics ($l=2$ for toroidal displacements $T$, $T_z$ and
$l=3$ for poloidal and radial displacements $Q$, $Q_z$, $\xi_{{\rm b} r}$, $z_r$) are plotted.
Bold lines denote Lagrangian displacements for baryons, and thin lines denote superfluid displacements.
Critical temperatures are constant throughout the core,
$T_{\rm cn} = 6 \times 10^8~{\rm K}$, $T_{\rm cp} = 5 \times 10^9~{\rm K}$.
Vertical dots show the crust-core interface.
}
\label{fig:eigenfuncs:i35}
\end{figure}

\begin{figure}
\centering
\begin{minipage}{.48\textwidth}
  \centering
  \includegraphics[width=1.\linewidth]{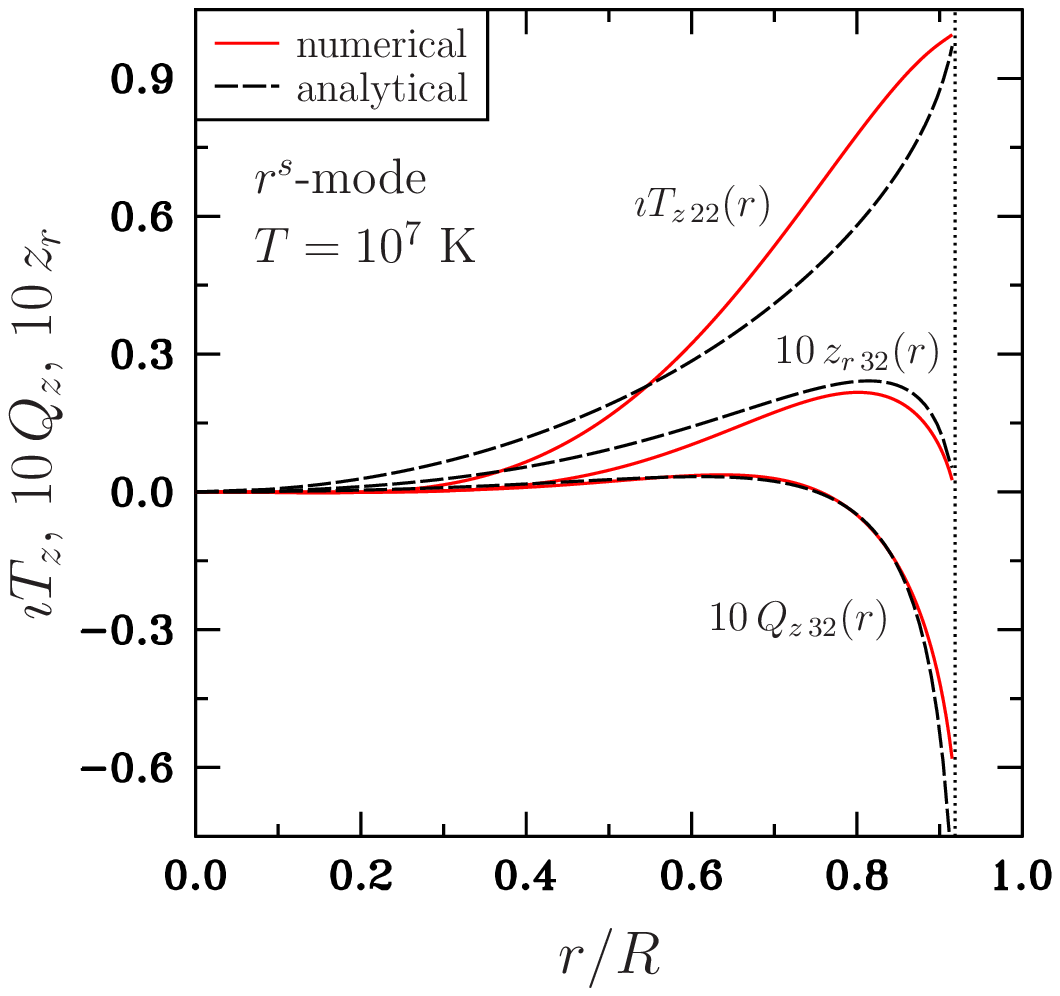}
\end{minipage}
\begin{minipage}{.48\textwidth}
  \centering
  \includegraphics[width=1.\linewidth]{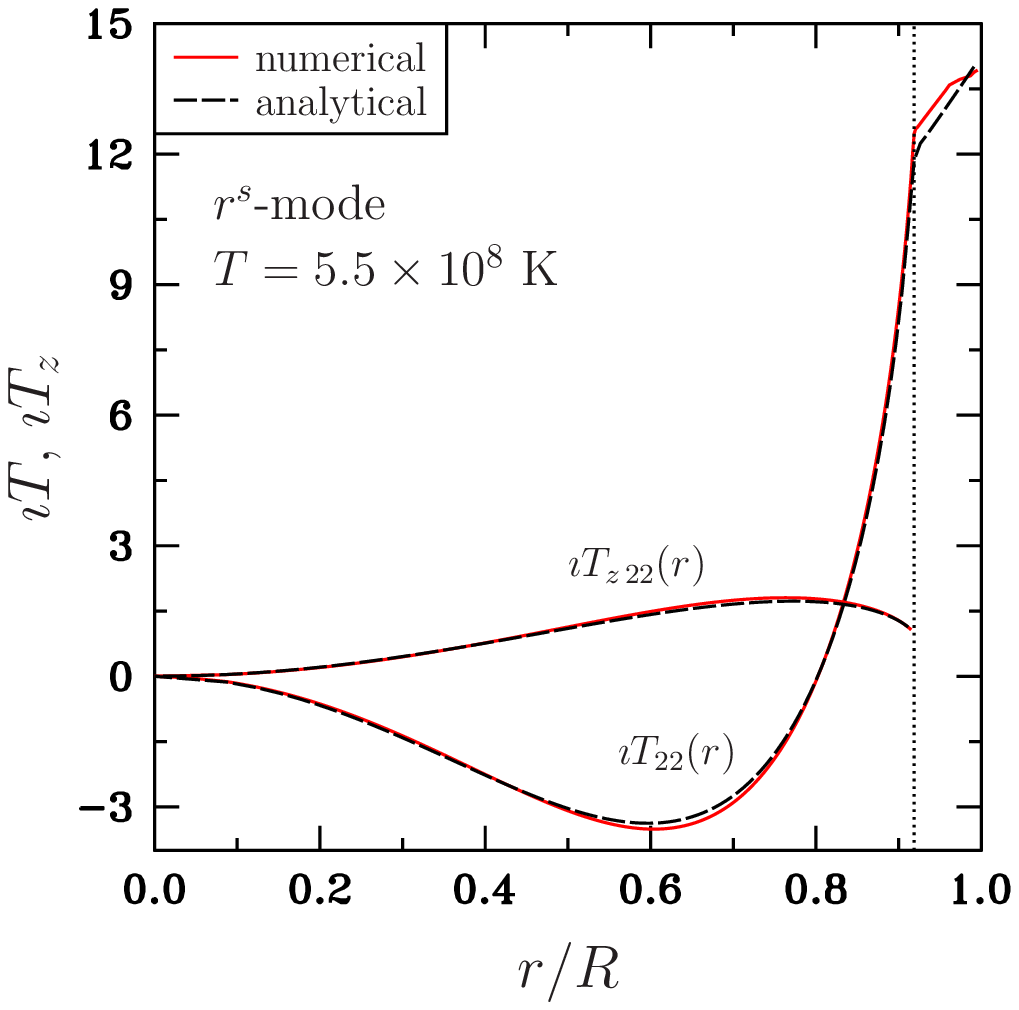}
\end{minipage}
\caption{
Dominant eigenfunctions for the $m=2$ superfluid $r$-mode,
obtained by numerical calculations (solid lines) and by the approximate analytical method (dashed lines).
Left panel displays superfluid displacements
$T_{z\,22}$, $Q_{z\,32}$ (multiplied by $10$), and $z_{r\,32}$ (multiplied by $10$) for $T = 10^7~{\rm K}$.
Right panel displays toroidal displacements $T_{z\,22}$ and $T_{22}$ for $T = 5.5 \times 10^8~{\rm K}$.
Critical temperatures are constant throughout the core,
$T_{\rm cn} = 6 \times 10^8~{\rm K}$, $T_{\rm cp} = 5 \times 10^9~{\rm K}$.
Vertical dots show the crust-core interface.
}
\label{fig:eigenfuncs:is1}
\end{figure}

\section{Results for the spectrum}
\label{sec:numerical}

In our numerical calculations we adopt the parametrization of \cite{hh99} 
of APR equation of state (\citealt{apr98}) 
for the NS core,
and the equation of state BSk20 \citep{pfcpg13} 
for the crust. All calculations are performed for an NS with mass $M = 1.4 M_\odot$
and radius $R = 12.18~{\rm km}$.
To calculate the unperturbed model of a star,
we use the Tolman-Oppenheimer-Volkoff equations.\footnote{
Note that here we, as \cite{kg17},
use relativistic background NS model and relativistic EOS,
but Newtonian oscillation equations.
One should bear in mind that this inconsistency may, in principle,
affect the results of our calculations.
}
We also assume that the baryon critical temperatures
are constant throughout the core:
$T_{\rm cn} = 6 \times 10^8~{\rm K}$,
$T_{\rm cp} = 5 \times 10^9~{\rm K}$
in all figures except for Fig.~\ref{fig:spectrumnpe-lowTcp},
where we set
$T_{\rm cn} = 5 \times 10^9~{\rm K}$,
$T_{\rm cp} = 5 \times 10^8~{\rm K}$.
The entrainment matrix $Y_{\rm ik}$
is calculated in a way similar to how it was done in \cite{kg11}.
The elements of this matrix are shown in Fig.~\ref{fig:Yik}
as functions of radial coordinate $r/R$ at temperature $T = 10^7~{\rm K}$ (left panel)
and as functions of temperature $T$ at fixed $r/R = 0.6$ (right panel).

The key ingredients of the scenario proposed by \cite{gck14a,gck14b} 
are the avoided crossings of inertial modes with the $m=2$ normal $r$-mode.
We calculated only the modes that could interact with this mode
i.e. the odd ($l_0 - |m| = 1$, $l_0 - |m| = 3$, and $l_0 - |m| = 5$)
$m=2$ inertial modes.\footnote{
	The odd modes with higher $l_0$, which also can interact with the $r^o$-mode,
	are harder to calculate numerically, so in the present study we focus
	only on the modes with low $l_0$.
}
We do not look for the modes with eigenfrequencies that are too far from the $r$-mode frequency
($\sigma_0 = 2/(m+1) = 2/3$ in the frame rotating with the star).
In order to solve the oscillation equations \eqref{eq:cont-b}--\eqref{eq:sfl-phi} numerically, we disregard all the terms with
$l > |m| + 2 k_{\rm max} - 1$ in the Legendre polynomial expansion.
We set $k_{\rm max} = 3$ to calculate $l_0 - |m| = 3$ and $l_0 - |m| = 5$ modes.
This value allows us to reproduce the results of \cite{yl00} for
the $l_0 - |m| = 3$ inertial modes within the accuracy of $0.2\%$.
For r-modes ($l_0 - |m| = 1$), for which only $l=m$ and $l=m+1$ harmonics are significant, we set $k_{\rm max} = 2$.

In Fig.~\ref{fig:spectrumnpe} we show the spectrum for
$l_0 - |m| = 1$, $l_0 - |m| = 3$, and $l_0 - |m| = 5$ inertial modes.
Dashed lines denote the i-modes calculated without entrainment ($Y_{\rm np} = 0$),
solid lines -- with entrainment. 
The bold line is the normal $m=2$ $r^o$-mode, $\sigma_0 = 2/3$
(note that if $Y_{\rm np} = 0$ the superfluid r-mode has the same frequency in the lowest order in $\Omega$,
see \citealt{ac01,ly03,agh09,kg17}). The dot-dashed line denotes the superfluid r-mode calculated analytically
via the approximate method described in Section~\ref{sec:r-mode}.
This approximate method accounts for the first-order terms in 
small parameter $\Delta h \equiv h_1/h - 1$
and thus it is accurate up to the terms $\sim \Delta h^2$.
In Fig.~\ref{fig:h1h} we plot the ratio $h_1(r)/h(r)$ for our NS model
at $T = 10^7~{\rm K}$ (dashed line)
and $T = 5.5 \times 10^8~{\rm K}$ (dot-dashed line).
As one can see from this Figure, $\Delta h$ at low temperatures 
is much larger than at temperatures close to $T_{\rm cn}$:
$\Delta h \gtrsim 0.2$ at $T = 10^7~{\rm K}$,
$\Delta h \lesssim 0.02$ at $T = 5.5 \times 10^8~{\rm K}$.
Therefore one can expect that at $T \to T_{\rm cn}$
the $r^s$-mode frequency will approach the value $\sigma_0 = 2/(m+1)$,
and the accuracy of our analytical method will increase.
Indeed, this conclusion is confirmed in Fig.~\ref{fig:spectrumnpe}.
One can see that even at low temperatures, when the frequency of $r^s$-mode
significantly differs from the normal $r^o$-mode frequency,
the two methods of calculating the frequency of $r^s$-mode give the same result
within the accuracy of $1\%$.
For example, at $T=10^7~{\rm K}$
numerical calculation for $r^s$-mode yields $\sigma_0 = 0.8452$,
while the analytical result is $\sigma_0 = 0.8476$.
At higher temperatures $T > 3 \times 10^8~{\rm K}$
the relative difference between numerical and analytical results does not exceed $0.01\%$.
Note, however, that the approximate method
does not provide such a good accuracy for eigenfunctions
(see discussion of Fig.~\ref{fig:eigenfuncs:is1} below).

The spectrum of inertial modes exhibits some interesting features:
(i) At low temperatures in the absence of entrainment the frequencies of the normal and superfluid i-modes 
almost coincide.
(ii) While the entrainment effect almost does not affect the normal inertial modes ($i^o$-modes),
it noticeably pushes the frequencies of superfluid modes up. We discuss this behavior in Section~\ref{sec:wkb}.
In order to illustrate the transition from the case $Y_{\rm np} \neq 0$ to $Y_{\rm np} = 0$,
we show in Fig.~\ref{fig:spectrumnpe-lowTcp} the spectrum for the same modes,
but employing another (non-realistic) ordering of critical temperatures:
$T_{\rm cp} = 5 \times 10^8~{\rm K}$, $T_{\rm cn} = 5 \times 10^9~{\rm K}$.
The absolute value of $Y_{\rm np}$ decreases as the temperature goes up,
and vanishes at $T = T_{\rm cp}$, when all protons become non-superconducting.
(iii) 
In contrast to the normal $r^o$-mode, which has the same frequency in the whole temperature range in both superfluid and non-superfluid NSs,
the frequencies of the normal $i^o$-modes
at temperatures close to $T_{\rm cn}$ do not remain constant but go down.
This is not surprising, since non-superfluid stratified $npe$-matter of NS cores does not support inertial modes.
Superfluid $i^s$-mode frequencies, in contrast, grow at $T \to T_{\rm cn}$.
In order to explain all these features, in Section~\ref{sec:wkb} we analyze a dispersion relation for inertial modes
in short-wavelength limit (see equations~\ref{eq:wkb-g-1}-\ref{eq:wkb-y-2} and their discussion).

In Fig.~\ref{fig:spectrumnpe} one can see avoided crossings of the $l_0 - |m| = 3$ normal $i^o$-mode
and its neighbouring $l_0 - |m| = 5$ superfluid $i^s$-mode (see solid lines)
at $T\sim 10^8~{\rm K}$ and at $T\sim 5.5 \times 10^8~{\rm K}$:
normal mode transforms into superfluid, and vice versa.
We did not find any avoided crossing between superfluid $i^s$-modes
and the normal $r^o$-mode, i.e. did not find the stability peaks for the scenario proposed by \cite{gck14a,gck14b}.
However, there should be an interaction between the normal and superfluid r-modes
at $T \to T_{\rm cn}$, which can stabilize normal $r^o$-mode, and result in the formation of the stability peak at $T = T_{\rm cn}$.
Notice also that at low temperatures the frequency of $l_0 - |m| = 3$ superfluid $i^s$-mode
is rather close to the frequency of the normal $r^o$-mode,
which may also lead to stabilizing interaction of modes at low temperatures.
To work out these interactions,
one has to go beyond the leading-order calculations in $\Omega$,
which is out of the scope of the present paper.
One may expect to find avoided crossings of $r^o$-mode with another inertial modes
under more realistic assumptions about the NS composition,
and/or for modes with larger $l_0$.
In the future work we are going to calculate the spectrum using realistic critical temperature profiles
and accounting for muons, which play very important role in defining oscillation spectrum
\citep{kg17}.

In order to illustrate the properties of inertial modes, we also plot their eigenfunctions.
Fig.~\ref{fig:eigenfuncs:i3} shows eigenfunctions for $2 \leq l \leq 7$ harmonics
of toroidal, poloidal and radial displacements for $l_0 - |m| = 3$ $i^o$-mode at $T = 10^7~{\rm K}$.
One can see that the baryon displacements (bold lines)
are larger than the superfluid ones (thin lines).
We also see that each of the dominant eigenfunctions
$T_{22}(r)$, $Q_{32}(r)$ and $\xi_{{\rm b}r\,32}(r)$ have one radial node,\footnote{
Following \cite{yl00}, we include the node at the stellar surface in the count of nodes.
}
in accordance with Table~3 in the paper by \cite{yl00}.

The lowest harmonics for $l_0 - |m| = 3$ and $l_0 - |m| = 5$ normal and superfluid $i$-modes
are plotted in Fig.~\ref{fig:eigenfuncs:i35}.
Eigenfunctions are normalized so that $\imath T_{22}(R_{cc}) = 1$ for $i^o$-modes,
and $\imath T_{z\,22}(R_{cc}) = 1$ for $i^s$-modes.
One can clearly see the key difference between the normal and superfluid modes:
for normal modes (top left and top right panel) baryon displacements (bold lines)
are comparable with (and even larger than) the superfluid ones (thin lines),
while for superfluid modes (bottom left and bottom right panel) superfluid displacements dominate.
It is also interesting that toroidal displacements are larger than poloidal for all considered modes.
The lowest-order dominant eigenfunctions
($T_{22}$, $Q_{32}$ and $\xi_{{\rm b}r\,32}$ for $i^o$-modes,
$T_{z\,22}$, $Q_{z\,32}$ and $z_{r\,32}$ for $i^s$-modes)
have one radial node for the case $l_0 - |m| = 3$ (left)
and two nodes for $l_0 - |m| = 5$ -- again, as expected from Table~3 in the paper by \cite{yl00}.
Note that, e.g., $T_{22}(r)$ for the $l_0 - |m| = 3$ $i^s$-mode
has more nodes than $T_{z\,22}(r)$ and cannot be used to determine the value of $l_0$.

We also compare eigenfunctions obtained via numerical calculations and via analytical method from Section~\ref{sec:r-mode}
for the superfluid $r^s$-mode
at low ($T = 10^7~{\rm K}$) and high ($T = 5.5 \times 10^8~{\rm K}$) temperatures.
The results are shown in Fig.~\ref{fig:eigenfuncs:is1}.
Solid lines represent the numerical results, dashed lines show analytical results.
For $T = 10^7~{\rm K}$ (left panel) we plotted only the superfluid displacements $T_{z\,22}(r)$, $Q_{z\,32}$ and $z_{r\,32}$ 
because they are much larger than the baryon ones.
Since the toroidal component is dominating, we multiplied $Q_{z\,32}$ and $z_{r\,32}$ by $10$ in order to make them visible.
We see that the numerical and analytical methods lead to qualitatively similar, but quantitatively different result:
while the eigenfrequencies coincide within the accuracy of $1\%$, eigenfunctions differ significantly.
At $T = 5.5 \times 10^8~{\rm K}$ (right panel), where entrainment effect is small (see Fig.~\ref{fig:h1h})
and the $r^s$-mode frequency is close to that of $r^o$-mode,
analytical method becomes more accurate. Indeed, we see that in this case the eigenfunctions coincide much better,
within the accuracy of $10\%$.
Here we plotted the superfluid and baryon toroidal displacements, $T_{z\,22}(r)$ and $T_{22}(r)$,
since at this temperature they are much larger than all other eigenfunctions.
The large value of $T_{22}(r)$ indicates a possible interaction with the normal $r$-mode.
From the analysis of Figs.~\ref{fig:spectrumnpe},~\ref{fig:spectrumnpe-lowTcp} and \ref{fig:eigenfuncs:is1}
we conclude that the approximate method of calculating the superfluid $r$-mode
gives a very good accuracy (better than $1\%$) for eigenfrequencies, but one has to keep in mind that
the eigenfunctions are calculated much less accurately, especially when entrainment effect is large (e.g., at low temperatures).

\section{Superfluid r-mode in the limit of small entrainment}
\label{sec:r-mode}

In this section we provide an approximate method that allows one to calculate the superfluid $r$-mode
in $npe$ NS analytically in the lowest order in $\Omega$, accounting for entrainment effect.
If there is no entrainment ($Y_{\rm np} = 0$ and thus $h_1 = h$, see equation~\ref{eq:h}), then for a given $m$ there exist
two purely toroidal rotational modes, the normal $r$-mode and the superfluid $r$-mode,
both having (to the lowest order in $\Omega/\Omega_0$) the same frequency $\sigma_0 = 2/(m+1)$
\citep{ac01,ly03,agh09,kg17}.
If the entrainment effect is present,
then,
except for some degenerate cases when $h_1$ is equal to $h$,\footnote{
As follows from equation~\ref{eq:h},
the condition $h_1 = h$
is equivalent to
$\mu_{\rm p} Y_{\rm np} = - \mu_{\rm n} Y_{\rm np}^2 / Y_{\rm pp}$.
This condition can be, in principle, satisfied even if $Y_{\rm np} \neq 0$.
}
the superfluid $r$-mode
turns into a mixed poloidal-toroidal mode with different frequency.\footnote{
If $h_1(r)/h(r)$ is constant throughout the core, the mode has different frequency
but remains purely toroidal.
}
Assuming that the entrainment effect is small, one can develop a
perturbation theory in $\Delta h \equiv h_1/h - 1$,
and analytically find corrections to the eigenfrequency and eigenfunctions
for the superfluid $r$-mode in the first order in $\Delta h$.
This method is analogous to that of \cite{kg17}, who showed
that in $npe$ matter, $r$-modes can be calculated analytically
in the next-to-leading order in $\Omega$, but ignoring the entrainment effect (and thus assuming $h_1 = h$).
\cite{pbr81} used a similar approach for analysis of $r$-modes in non-superfluid stars.

Let us start with purely toroidal oscillations, described by system~\eqref{eq:cont-b}-\eqref{eq:sfl-phi}.
In that case radial displacements vanish ($\xi_{{\rm b} r} = z_r = 0$),
and the continuity equations \eqref{eq:cont-b} and \eqref{eq:cont-e} reduce to
\begin{gather}
\label{eq:cont-b-toroidal}
	\pd{}{\theta} \sin\theta \xi_{{\rm b} \theta}^0
	+ \imath m \xi_{{\rm b} \phi}^0 = 0
,\\
\label{eq:cont-e-toroidal}
	\pd{}{\theta} \sin\theta z_\theta^0 
	+ \imath m z_\phi^0 
	= 0
.
\end{gather}
$\theta$-components of Euler equation \eqref{eq:euler-th} and superfluid equation \eqref{eq:sfl-th} read
\begin{gather}
\label{eq:euler-th-toroidal}
	- \sigma_0 \xi_{{\rm b}\theta}^0 - 2 \imath \cos\theta \xi_{{\rm b}\phi}^0
	= \frac{1}{\imath m} \pd{}{\theta} 
	\sin\theta
	\left(
		- \sigma_0 \xi_{{\rm b}\phi}^0 + 2 \imath \xi_{{\rm b}\theta}^0 \cos\theta 
	\right)
,\\
\label{eq:sfl-th-toroidal}
	- \sigma_0 z_\theta^0 - 2 \imath \frac{h_1}{h} \cos\theta z_\phi^0
	= \frac{1}{\imath m} \pd{}{\theta} 
	\sin\theta
	\left(
		- \sigma_0 z_\phi^0 + 2 \imath \frac{h_1}{h} z_\theta^0 \cos\theta
	\right)
.
\end{gather}

The solution to the system of equations \eqref{eq:cont-b-toroidal} and \eqref{eq:euler-th-toroidal} is
\begin{gather}
\label{eq:r-mode-toroidal-normal}
	\sigma_0 = \frac{2m}{l (l+1)}
,\quad
	\xi_{{\rm b}\theta}^0 = \frac{\imath m}{{\rm sin}\theta} T_{lm}(r) P_l^m (\cos\theta)
,\quad
	\xi_{{\rm b}\phi}^0 =  - T_{lm}(r) \d{}{\theta}P_l^m (\cos\theta)
,
\end{gather}
which is the well-known normal $r$-mode.
Taking into account the $r$- and $\phi$-components of Euler equation
(equations \ref{eq:euler-r} and \ref{eq:euler-phi}),
one can check that only the solution with $l = m$ exists.

The second pair of equations, \eqref{eq:cont-e-toroidal} and \eqref{eq:sfl-th-toroidal}, describes the superfluid $r$-mode,
\begin{gather}
\label{eq:r-mode-toroidal-sfl}
	\sigma_0 = \frac{2m}{l (l+1)} \frac{h_1(r)}{h(r)}
,\quad
	z_\theta^0 = \frac{\imath m}{{\rm sin}\theta} T_{z\,lm}(r) P_l^m (\cos\theta)
,\quad
	z_\phi^0 =  - T_{z\,lm}(r) \d{}{\theta}P_l^m (\cos\theta)
.
\end{gather}
If $Y_{\rm np}=0$, then $h_1 = h$, and $\sigma_0 = \frac{2m}{l (l+1)}$ is the global oscillation frequency and the superfluid $r$-mode
is indeed purely toroidal (and, as for the normal $r$-mode, only $l=m$ solution exists).
However, if the entrainment effect is present, $h_1(r)/h(r)$ in general case varies throughout the star.
This means that the purely toroidal superfluid mode cannot exist, and an admixture
of poloidal component is required.

Now let us write down a perturbation theory in $\Delta h \equiv h_1/h - 1$.
Below we denote the zeroth-order in $\Delta h$ quantities with index $(0)$,
and the first-order in $\Delta h$ quantities -- with index $(1)$.

In this notation, the eigenfrequency $\sigma_0$ and the eigenfunctions can be expanded in Taylor series in $\Delta h$:
\begin{gather}
	\sigma_0 
	= \sigma_{0(0)}  + \sigma_{0(1)} + O(\Delta h^2)
	= \frac{2}{m+1} + \sigma_{0(1)}  + O(\Delta h^2)
,\\
	\xi_{{\rm b} r}^0 = \xi_{{\rm b} r}^{0(1)}  + O(\Delta h^2)
,\quad
	T = T^{(0)} + T^{(1)}   + O(\Delta h^2)
,\quad
	Q = Q^{(1)}  + O(\Delta h^2)
,\\
	z_r^0 = z_r^{0(1)}  + O(\Delta h^2)
,\quad
	T_z = T_z^{(0)} + T_z^{(1)}  + O(\Delta h^2)
,\quad
	Q_z = Q_z^{(1)} + O(\Delta h^2)
,\\
	\delta P = \delta P^{1(0)} + \delta P^{1(1)}  + O(\Delta h^2)
,\\
	\Delta \mu_{\rm e}^1 = \Delta \mu_{\rm e}^{1(0)} + \Delta \mu_{\rm e}^{1(1)}  + O(\Delta h^2)
.
\end{gather}
Since in absence of entrainment the superfluid $r$-mode is purely toroidal,
the radial and poloidal displacements in the zeroth order vanish,
$\xi_{{\rm b} r}^{0(0)} = z_r^{0(0)} = Q^{(0)} = Q_z^{(0)} = 0$.

\subsection{Zero-order solution}

In the zeroth order in $\Delta h$ (i.e. without the entrainment effect)
one has to find the eigenfrequency $\sigma_{0(0)}$
and four eigenfunctions $T^{(0)}, T_z^{(0)}, \delta P^{1(0)}, \Delta\mu_{\rm e}^{1(0)}$.
As discussed above (see equations~\ref{eq:r-mode-toroidal-normal} and \ref{eq:r-mode-toroidal-sfl}),
the frequency equals to
\begin{gather}
	\sigma_{0(0)} = \frac{2}{m+1}
,
\end{gather}
and the toroidal displacements are proportional to the $l=m$ Legendre polynomial,
\begin{gather}
	T^{(0)} = T_{mm}^{(0)} (r) P_m^m (\cos\theta)
,\quad
	T_z^{(0)} = T_{zmm}^{(0)} (r) P_m^m (\cos\theta)
.
\end{gather}
One can find from equations~\eqref{eq:euler-r}, \eqref{eq:euler-phi}, \eqref{eq:sfl-r}, and \eqref{eq:sfl-phi}
that the perturbations $\delta P^{1(0)}$ and $\Delta\mu_{\rm e}^{1(0)}$ are proportional to the $l=m+1$ Legendre polynomial,
\begin{gather}
	\delta P^{1(0)}	= \delta P_{m+1,m}^{1(0)} (r) P_{m+1}^m (\cos\theta)
,\quad
	\Delta\mu_{\rm e}^{1(0)} = \Delta\mu_{{\rm e}\,m+1,m}^{1(0)} (r) P_{m+1}^m (\cos\theta)
,
\end{gather}
while the coefficients $\delta P_{m+1,m}^{1(0)} (r)$, $\Delta \mu_{{\rm e}\,m+1,m}^{1(0)} (r)$
are expressed through $T_{mm}^{(0)}(r)$ and $T_{zmm}^{(0)}(r)$, respectively:
\begin{gather}
	\delta P_{m+1,m}^{1(0)} (r)
	= \frac{\imath \sigma_{0(0)} (\sigma_{0(0)}-2)}{2m+1} w r T_{mm}^{(0)}(r)
,\\
	\Delta \mu_{{\rm e}\,m+1,m}^{1(0)} (r)
	= \frac{\imath \sigma_{0(0)} (\sigma_{0(0)}-2)}{2m+1} \frac{h}{c^2 n_{\rm e}} r T_{zmm}^{(0)}(r)
.
\end{gather}

After substituting the expressions for $\delta P^{1(0)}$ and $\Delta\mu_{\rm e}^{1(0)}$
into equations~\eqref{eq:euler-r} and \eqref{eq:sfl-r}, one 
can finally obtain the solution for $T_{zmm}^{(0)}(r)$ and $T_{mm}^{(0)}(r)$
\citep[see][Appendix B]{kg17},
\begin{gather}
\label{eq:Tzmm0}
	T_{zmm}^{(0)}(r) = C_1 \frac{n_{\rm e}(r)}{h(r)} r^m
,\\
\label{eq:Tmm0}
	T_{mm}^{(0)}(r) = r^m \left(C_0 +
			C_1 \int_{0}^{r}
				\frac{\mu_{\rm n} (r_1)}{c^4 w^2(r_1)}
				\d{P(r_1)}{r_1}
				\pd{n_{\rm b}}{\Delta\mu_{\rm e}}(r_1)
				{\rm d} r_1
		\right)
.
\end{gather}

The integration constants $C_0$ and $C_1$ have to be determined from the first-order equations.

\subsection{First-order solution}
To find the eigenfrequency correction $\sigma_{0(1)}$ and the constants $C_0$ and $C_1$,
it is sufficient to consider only the continuity equations \eqref{eq:cont-b}--\eqref{eq:cont-e}
as well as $\theta$-components of the Euler equation \eqref{eq:euler-th} and the superfluid equation \eqref{eq:sfl-th}.

$\theta$-component of the Euler equation reads, in first order in $\Delta h$
[i.e. ignoring quadratically small terms like $\sigma_{0(1)} \xi_{{\rm b}\theta}^{0(1)}$],
\begin{gather}
\label{eq:euler-th-11}
	- \sigma_{0(1)} \xi_{{\rm b}\theta}^{0(0)}
	- \sigma_{0(0)} \xi_{{\rm b}\theta}^{0(1)}
	- 2 \imath \cos\theta \xi_{{\rm b}\phi}^{0(1)}
	= \frac{1}{\imath m} \pd{}{\theta} 
	\sin\theta
	\left[
		- \sigma_{0(1)} \xi_{{\rm b}\phi}^{0(0)}
		- \sigma_{0(0)} \xi_{{\rm b}\phi}^{0(1)}
		+ 2 \imath \left( \xi_{{\rm b}r}^{0(1)} \sin\theta + \xi_{{\rm b}\theta}^{0(1)} \cos\theta \right) 
	\right]
.
\end{gather}
Substituting relations \eqref{eq:xib-tp}, \eqref{eq:lm-xib}, \eqref{eq:lm-Q}, \eqref{eq:lm-T}
into equation~\eqref{eq:euler-th-11} divided by $\sin\theta$
and equating coefficients at the terms proportional to $P_m^m$,
one can express $Q_{m+1,m}^{(1)} (r)$ through $\xi_{{\rm b} r\, m+1,m}^{0(1)} (r)$ and $T_{mm}^{(0)} (r)$.

Similarly, using
$\theta$-component of the superfluid equation,
\begin{multline}
\label{eq:sfl-th-11}
	- \sigma_{0(1)} z_\theta^{0(0)}
	- \sigma_{0(0)} z_\theta^{0(1)}
	- 2 \imath \Delta h \cos\theta z_\phi^{0(0)}
	- 2 \imath \cos\theta z_\phi^{0(1)}
\\
	= \frac{1}{\imath m} \pd{}{\theta} 
	\sin\theta
	\left[
		- \sigma_{0(1)} z_\phi^{0(0)}
		- \sigma_{0(0)} z_\phi^{0(1)}
		+ 2 \imath \left( z_r^{0(1)} \sin\theta + z_\theta^{0(1)} \cos\theta \right)
		+ 2 \imath \Delta h z_\theta^{0(0)} \cos\theta
	\right]
,
\end{multline}
one can obtain an algebraic relation between
$Q_{z \,m+1,m}^{(1)} (r)$, $z_{r\, m+1,m}^{0(1)} (r)$ and $T_{z \,mm}^{(0)} (r)$.

Now, taking the coefficient at $P_{m+1}^m$ in the continuity equation for baryons \eqref{eq:cont-b}
\begin{gather}
	\frac{1}{n_{\rm b}} \frac{1}{r^2}\pd{}{r}r^2 n_{\rm b} \xi_{{\rm b} r}^{0(1)}
	+ \frac{1}{r\sin\theta}
	\left[
		\pd{}{\theta}\sin\theta \pd{Q^{(1)}}{\theta}
		- \frac{m^2 Q^{(1)}}{\sin\theta}
	\right]
	= 0
,
\end{gather}
expressing $Q^{(1)}_{m+1,m}$ through $T^{(0)}_{mm}$ and $\xi_{{\rm b}r, {m+1,m}}^{0(1)}$,
and substituting expression for $T^{(0)}_{mm}$ \eqref{eq:Tmm0},
we get a first-order inhomogeneous ODE for $\xi_{{\rm b} r, m+1,m}$:
\begin{gather}
	\d{}{r} \xi_{{\rm b} r, m+1,m}^{0(1)}
	+ A(r) \xi_{{\rm b} r, m+1,m}^{0(1)}
	- \sigma_{0(1)} C_0 B_{10}(r)
	- \sigma_{0(1)} C_1 B_{11}(r)
	= 0
,
\end{gather}
where $A(r), B_{10}(r), B_{11}(r)$ are known functions of $r$.
The solution to this equation is
\begin{gather}
	\xi_{{\rm b} r, m+1,m}^{0(1)} (r)
	= H(r) \left[
			\xi_0
			+ \sigma_{0(1)} C_0 \int_{0}^{r} \frac{B_{10}(x)}{H(x)}{\rm d}x
			+ \sigma_{0(1)} C_1 \int_{0}^{r} \frac{B_{11}(x)}{H(x)}{\rm d}x
		\right]
,\quad
	H(r) \equiv \exp \left( -\int A(r) {\rm d}r \right)
		 = \frac{1}{n_{\rm b} (r) r^{m+3}} 
.
\end{gather}
Since
$\xi_{{\rm b} r, m+1,m}^{0(1)} (r)$ should be finite at $r \to 0$
the integration constant $\xi_0 = 0$.

Following the same procedure for electron continuity equation~\eqref{eq:cont-e},
we obtain the expression for $z_{r, m+1,m}^{0(1)} (r)$,
\begin{gather}
	z_{r, m+1,m}^{0(1)} (r)
	= H_z(r) \left[
			z_0
			+ C_1 \int_{r_{\rm sfl 1}}^{r} \frac{B_{z01}(x)}{H_z(x)}{\rm d}x
			+ \sigma_{0(1)} C_0 \int_{r_{\rm sfl 1}}^{r} \frac{B_{z10}(x)}{H_z(x)}{\rm d}x
			+ \sigma_{0(1)} C_1 \int_{r_{\rm sfl 1}}^{r} \frac{B_{z11}(x)}{H_z(x)}{\rm d}x
		\right]
,\\
	H_z(r) \equiv \exp \left( -\int A_z(r) {\rm d}r \right)
		 = \frac{1}{n_{\rm e} (r) r^{m+3}} 
,
\end{gather}
where $A_z(r), B_{z01}(r), B_{z10}(r), B_{z11}(r)$ are known functions of $r$,
and $r_{\rm sfl 1}$ is the inner boundary of superfluid region.
If $r_{\rm sfl 1} = 0$, then the integration constant $z_0 = 0$,
because $z_r$ should be finite at $r = 0$;
otherwise $z_0$ is still zero
because of the boundary condition
$z_r = 0$ at the boundary of superfluid region.

The finiteness of $\xi_{{\rm b} r}$ at the stellar surface $r=R$
and vanishing of $z_r$ at the outer superfluid boundary $r = r_{\rm sfl 2}$
imply
\begin{gather}
\label{eq:r-mode-const-1}
	\sigma_{0(1)} C_0 \int_{0}^{R} \frac{B_{10}(x)}{H(x)}{\rm d}x
	+ \sigma_{0(1)} C_1 \int_{0}^{R} \frac{B_{11}(x)}{H(x)}{\rm d}x
	= 0
,\\
\label{eq:r-mode-const-2}
	C_1 \int_{r_{\rm sfl 1}}^{r_{\rm sfl 2}} \frac{B_{z01}(x)}{H_z(x)}{\rm d}x
	+ \sigma_{0(1)} C_0 \int_{r_{\rm sfl 1}}^{r_{\rm sfl 2}} \frac{B_{z10}(x)}{H_z(x)}{\rm d}x
	+ \sigma_{0(1)} C_1 \int_{r_{\rm sfl 1}}^{r_{\rm sfl 2}} \frac{B_{z11}(x)}{H_z(x)}{\rm d}x
	= 0
.
\end{gather}
The system \eqref{eq:r-mode-const-1}--\eqref{eq:r-mode-const-2}
has two independent solutions.\footnote{The constant $C_0$ can be set to arbitrary value (e.g. $C_0 = 1$)
by choosing an appropriate normalization for eigenfunctions.}
The first solution is $\sigma_{0(1)} = C_1 = 0$; it is the normal $r$-mode,
\begin{gather}
	\sigma_0 = \sigma_{0(0)} = \frac{2}{m+1}
,\quad
	T_{mm}^{(0)} = C_0 r^m
,\quad
	T_{zmm}^{(0)} = 0
.
\end{gather}
The second one, having $\sigma_{0(1)} \neq 0$ and $C_1 \neq 0$, is the superfluid $r$-mode.


We compared this analytical solution with the numerical one (see Section~\ref{sec:numerical}) and found
that the difference between the eigenfrequencies calculated with these two approaches
does not exceed $1\%$ even at low temperatures,
where $\Delta h$ is relatively large, $\Delta h \sim 0.2-0.25$.

The next possible step would be to combine two approximate methods and calculate $r$-mode eigenfrequencies
accounting for both entrainment and next-to-leading-order terms in $\Omega$.
Entrainment may significantly shift the resonance temperatures,
where an avoided crossing of normal and superfluid $r$-mode occurs,
and thus affect the shape of $r$-mode instability window.

\section{Short-wavelength analysis for inertial modes}
\label{sec:wkb}

In this section we obtain a dispersion relation for inertial modes in superfluid $npe$ matter
and analyze it in different limiting cases in order to explain behaviour of the modes at low and high temperatures.

Let us find the dispersion relation for inertial modes defined by equations \eqref{eq:cont-b}--\eqref{eq:sfl-phi}
in the short-wavelength limit, in which derivatives of any perturbation $\delta A$ can be replaced
as $\pd{}{\v{r}} \delta A \to - \imath \v{k} \delta A$,
where $\v{k}$ is the wave vector.

Continuity equation for baryons \eqref{eq:cont-b} in this limit reads:
\begin{gather}
\label{eq:wkb:cont-b}
	- \imath \v{k} \v{\xi}_{\rm b}^0
	= 0
.
\end{gather}
Here we omitted the term
$\xi_{{\rm b}r}^0 {\rm d}(\ln n_{\rm b}) / {\rm d}r$
in comparison to $- \imath \v{k} \v{\xi}_{\rm b}^0$,
since the wavelength is assumed to be much smaller
than the density scale height, $k^{-1} \ll \abs{{\rm d} \ln n_{\rm b} / {\rm d}r}^{-1}$.

Subtracting continuity equation for electrons \eqref{eq:cont-e} from equation~\eqref{eq:cont-b}, one obtains
\begin{gather}
\label{eq:wkb:cont-e}
	- \imath \v{k} \v{z}^0 - \xi_{{\rm b}r}^0 \d{\ln x_{\rm e}}{r} = 0
,\quad x_{\rm e} \equiv \frac{n_{\rm e}}{n_{\rm b}}
.
\end{gather}

Euler equation \eqref{eq:euler-r}--\eqref{eq:euler-phi} 
and superfluid equation \eqref{eq:sfl-r}--\eqref{eq:sfl-phi} read:
\begin{gather}
\label{eq:wkb:euler-r}
	- \sigma_0^2 \xi_{{\rm b}r}^0 - 2 \imath \sigma_0 \sin\theta \xi_{{\rm b}\phi}^0
	= \imath k_r \frac{\delta P^1}{w}
	+ \frac{\mu_{\rm n}}{w^2 c^2}
	  		\pd{n_{\rm b}}{\Delta\mu_{\rm e}} \Delta\mu_{\rm e}^1
	  	\d{P}{r}
,\\
\label{eq:wkb:euler-th}
	- \sigma_0^2 \xi_{{\rm b}\theta}^0 - 2 \imath \sigma_0 \cos\theta \xi_{{\rm b}\phi}^0
	= \imath k_\theta \frac{\delta P^1}{w}
,\\
\label{eq:wkb:euler-phi}
	- \sigma_0^2 \xi_{{\rm b}\phi}^0 + 2 \imath \sigma_0
		\left( \xi_{{\rm b}r}^0 \sin\theta + \xi_{{\rm b}\theta}^0 \cos\theta \right) 
	= \imath k_\phi \frac{\delta P^1}{w}
,\\
\label{eq:wkb:sfl-r}
	- \sigma_0^2 z_r^0 - 2 \imath \frac{h_1}{h} \sigma_0 \sin\theta z_\phi^0
	= \imath k_r \frac{c^2 n_{\rm e}}{h} \Delta \mu_{\rm e}^1
,\\
\label{eq:wkb:sfl-th}
	- \sigma_0^2 z_\theta^0 - 2 \imath \sigma_0 \frac{h_1}{h} \cos\theta z_\phi^0
	= \imath k_\theta \frac{c^2 n_{\rm e}}{h} \Delta \mu_{\rm e}^1
,\\
\label{eq:wkb:sfl-phi}
	- \sigma_0^2 z_\phi^0 + 2 \imath \frac{h_1}{h} \sigma_0 \left( z_r^0 \sin\theta + z_\theta^0 \cos\theta \right)
		= \imath k_\phi \frac{c^2 n_{\rm e}}{h} \Delta \mu_{\rm e}^1
.
\end{gather}

Equations \eqref{eq:wkb:cont-b}--\eqref{eq:wkb:sfl-phi} can be written in a form
$\mathsf{\mathbf{A}} \cdot \v{x} = 0$, where
$\v{x} = \left(\delta P^1, \Delta\mu_{\rm e}^1,
	\xi_{{\rm b} r}^0, \xi_{{\rm b} \theta}^0, \xi_{{\rm b} \phi}^0,
	z_r^0, z_\theta^0, z_\phi^0
	\right)$,
and $\mathsf{\mathbf{A}}$ is a $8 \times 8$ matrix.
Dispersion relation between the frequency $\sigma = \sigma_0 \Omega$
and the wave vector $\v{k}$ can be found by solving the equation $\det \mathsf{\mathbf{A}} = 0$,
which reduces to a biquadratic equation in $\sigma_0$,
\begin{gather}
	A \sigma_0^4 + B \sigma_0^2 + C = 0
,
\end{gather}
where the coefficients $A,B,C$ are defined as
\begin{gather}
\label{eq:A}
	A = k^4 - \frac{y}{n_{\rm e}^2} \pd{n_{\rm b}}{\Delta\mu_{\rm e}} \d{P}{r} \d{x_{\rm e}}{r} \left(k^2-k_r^2\right)
,\\
\label{eq:B}
	B =
	- 4 k^2 \frac{(\v{\Omega k})^2}{\Omega ^2} \left[\left(\frac{h_1}{h}\right)^2 + 1 \right] 
	+ 4 \left(\frac{h_1}{h}\right)^2
		\frac{y}{n_{\rm e}^2}
		\pd{n_{\rm b}}{\Delta\mu_{\rm e}} \d{P}{r} \d{x_{\rm e}}{r}
		(k^2-k_r^2)
,\\
\label{eq:C}
	C =  16 \left(\frac{h_1}{h}\right)^2 \frac{(\v{\Omega k})^4}{\Omega^4}
.
\end{gather}

To get further insight into the problem, let us introduce the following
quantities: equilibrium speed of sound $c_{\rm eq}$,
adiabatic speed of sound $c_{\rm fr}$
and the coupling parameter $s$, which are defined as
\begin{gather}
	c_{\rm eq}^2 \equiv c^2 \frac{{\rm d}P /{\rm d}r}{\mu_{\rm n} {\rm d}n_{\rm b} /{\rm d}r}
,\quad
	c_{\rm fr}^2 \equiv c^2 \frac{1}{\mu_{\rm n}} \left(\pd{P}{n_{\rm b}}\right)_{x_{\rm e}}
,\quad
	s \equiv \frac{n_{\rm e} \left(\partial{P} / \partial{n_{\rm e}}\right)_{n_{\rm b}} }
			 {n_{\rm b} \left(\partial{P} / \partial{n_{\rm b}}\right)_{x_{\rm e}} }
.
\end{gather}
The derivative $\pd{n_{\rm b}}{\Delta\mu_{\rm e}}$ can be expressed in terms of these variables as
\begin{gather}
	\left( \pd{n_{\rm b}}{\Delta\mu_{\rm e}} \right)_P
	= \frac{n_{\rm b} n_{\rm e}}{s w}
		\left( \frac{1}{c_{\rm eq}^2} - \frac{1}{c_{\rm fr}^2}\right)
.
\end{gather}
Also, using the hydrostatic equilibrium condition, we are able to express
gradients of equilibrium quantities ($P$ and $x_{\rm e}$)
through gravitational acceleration $g$,
\begin{gather}
	\d{P}{r} = - w g
,\quad
	\d{x_{\rm e}}{r} = g \frac{x_e}{s} \left( \frac{1}{c_{\rm eq}^2} - \frac{1}{c_{\rm fr}^2}\right)
.
\end{gather}

Further, let us introduce
the Brunt-V$\ddot{\rm a}$is$\ddot{\rm a}$l$\ddot{\rm a}$ frequency
$\mathcal{N} \equiv g \left( 1/c_{\rm eq}^2 - 1/c_{\rm fr}^2 \right)^{1/2}$,
which enters the dispersion relation for $g$-modes
\citep{rg92},
and the `superfluid' speed $c_{\rm SFL} \equiv s c_{\rm eq} c_{\rm fr} / \sqrt{y (c_{\rm fr}^2 - c_{\rm eq}^2 )}$.
The latter quantity in the limit $y \to \infty$ (or, equivalently, $T \to T_{\rm cn}$),
equals to the superfluid speed of sound \citep[see e.g.][]{ac01,ga06}.

Using the above definitions, the coefficients $A$ and $B$ from equations~\eqref{eq:A}--\eqref{eq:B}
can be rewritten in the form
\begin{gather}
\label{eq:A2}
	A = k^4 + \frac{\mathcal{N}^2 (k^2 - k_r^2)}{c_{\rm SFL}^2}
,\\
\label{eq:B2}
	B =
	- 4 k^2 \frac{(\v{\Omega k})^2}{\Omega ^2} \left[\left(\frac{h_1}{h}\right)^2 + 1 \right] 
	- 4 \left(\frac{h_1}{h}\right)^2 \frac{\mathcal{N}^2 (k^2 - k_r^2)}{c_{\rm SFL}^2}
.
\end{gather}
Now, if we substitute $A$, $B$ and $C$ into a dispersion relation $\sigma^2 = \Omega^2 (-B \pm \sqrt{B^2-4AC})/(2A)$,
the result will be rather lengthy.
To make it more clear, let us note
that at low temperatures the ratio $\mathcal{N} / (c_{\rm SFL}k)$ is small, $\mathcal{N} / (c_{\rm SFL}k) \ll 1$.
For example, for a wavenumber $k = 10^{-5}~{\rm cm}^{-1}$
and a NS model used in Section~\ref{sec:numerical}
at a distance from the center $r = R/2$ and temperature $T = 10^7~{\rm K}$
this ratio equals $\mathcal{N} / (c_{\rm SFL}k) = 0.013$.
\footnote{
We remind the reader that the wavenumber $k$ is assumed to be large;
in particular, it is much greater than the inverse stellar radius,
$k \gg 1/R = 8.2\times 10^{-7}~{\rm cm}^{-1}$.}

In this limit the dispersion relation has the following form,
\begin{gather}
\label{eq:wkb-g-1}
	\sigma^2
	= 4 \frac{(\v{\Omega}\v{k})^2}{k^2}
		- 4 \frac{\mathcal{N}^2}{c_{\rm SFL}^2 k^2}
			\frac{(k^2-k_r^2)}{k^4} 
			\frac{\left(\frac{h_1}{h}\right)^2 \Omega^2 k^2-(\v{\Omega}\v{k})^2}
				{\left(\frac{h_1}{h}\right)^2-1}
		+ O\left(\frac{\mathcal{N}^4}{c_{\rm SFL}^4 k^4}\right)
,\\
\label{eq:wkb-g-2}
	\sigma^2
	= 4 \left(\frac{h_1}{h}\right)^2 \frac{(\v{\Omega}\v{k})^2}{k^2}
		+ 4 \frac{\mathcal{N}^2}{c_{\rm SFL}^2 k^2}
		\frac{(k^2-k_r^2)}{k^4} 
		\left(\frac{h_1}{h}\right)^4
		\frac{\Omega^2 k^2 - (\v{\Omega}\v{k})^2}
		   {\left(\frac{h_1}{h}\right)^2 - 1}
		+ O\left(\frac{\mathcal{N}^4}{c_{\rm SFL}^4 k^4}\right)
.
\end{gather}

The first relation \eqref{eq:wkb-g-1} describes normal $i^o$-modes.
In barotropic matter, where $\mathcal{N} = 0$, they have a standard dispersion relation,
$\sigma^2 = 4 (\v{\Omega}\v{k})^2 / k^2$ \citep{ll87}.
The second relation \eqref{eq:wkb-g-2} describes superfluid $i^s$-modes,
for which the leading term differs from that of $i^o$-modes
by the factor $(h_1/h)^2$.
One can conclude that, as the entrainment effect decreases,
for a given superfluid $i^s$-mode,
its frequency approaches the frequency of its normal ($i^o$-mode) counterpart.
Indeed, we observe such behaviour for $i^s$-modes in Fig.~\ref{fig:spectrumnpe-lowTcp} at $T < T_{\rm cp}$.
Note, however, that the relations~\eqref{eq:wkb-g-1} and \eqref{eq:wkb-g-2}
are invalid if there is no entrainment at all ($Y_{\rm np} = 0$ and therefore $h_1 = h$),
since they contain terms that are proportional to $\left[ (h_1/h)^2-1 \right]^{-1}$.
The asymptotic expansion for the case $h_1 = h$ reads
\begin{gather}
\label{eq:wkb-g-3}
	\sigma^2
	= 4 \frac{\left(\v{\Omega} \v{k}\right)^2}{k^2}	
	  \pm 4 \frac{\mathcal{N}}{c_{\rm SFL} k}
	  	\frac{\sqrt{(k^2 - k_r^2)}}{k^3}
	  	\abs{\v{\Omega} \v{k}}  \sqrt{\Omega^2 k^2 - (\v{\Omega} \v{k})^2}
	+ O \left( \frac{\mathcal{N}^2}{c_{\rm SFL}^2 k^2} \right)
,
\end{gather}
where the `$+$' sign refers to $i^s$-mode, and the `$-$` sign refers to the $i^o$-mode.
The frequencies of normal and superfluid $i$-modes coincide up to the 
terms proportional to $\mathcal{N}/(c_{\rm SFL}k) \ll 1$;
indeed, one can see in Fig.~\ref{fig:spectrumnpe},
that the frequencies of $l_0 - m = 3$ and $l_0 - m = 5$ $i^s$-modes
in the case $Y_{\rm np}=0$ are very close to the corresponding $i^o$-modes at
low temperatures $T \ll T_{\rm cn}$.
Fig.~\ref{fig:spectrumnpe-lowTcp} also illustrates this point:
when protons become non-superfluid (at $T > T_{\rm cp}$), the entrainment effect vanishes
and, while the temperature is still much less than $T_{\rm cn}$, $i^s$-modes and $i^o$-modes with the same $l_0$ are close to each other.

Now let us examine the behaviour of inertial modes in the limit of high temperatures $T \to T_{\rm cn}$.
Since the quantity $y$ in this limit tends to infinity, $y \to \infty$, the `superfluid` speed tends to zero, $c_{\rm SFL} \to 0$.
The Brunt-V$\ddot{\rm a}$is$\ddot{\rm a}$l$\ddot{\rm a}$ frequency, on the other hand, does not depend on temperature.
Thus, the ratio $\mathcal{N}/(c_{\rm SFL}k)$ can approach arbitrarily large values,
and asymptotic expansions \eqref{eq:wkb-g-1}-\eqref{eq:wkb-g-3} are invalid.
In this case one should consider the opposite limit, $c_{\rm SFL}k / \mathcal{N} \ll 1$,
in which the asymptotic expansion for the dispersion relation reads
\begin{gather}
\label{eq:wkb-y-1}
	\sigma^2
	= 4 \frac{c_{\rm SFL}^2 k^2}{\mathcal{N}^2}
		\frac{(\v{\Omega} \v{k})^4}
			{\Omega^2 k^2 \left(k^2-k_r^2 \right)}
		+ O\left(\frac{c_{\rm SFL}^4 k^4}{\mathcal{N}^4}\right)
,\\
\label{eq:wkb-y-2}
	\sigma^2		 
	= 4 \left(\frac{h_1}{h}\right)^2 \Omega^2
	- 4 \frac{c_{\rm SFL}^2 k^2}{\mathcal{N}^2}
		\frac{\left[\Omega^2 k^2-(\v{\Omega}\v{k})^2\right]
				\left[(\frac{h_1}{h})^2 \Omega^2 k^2 - (\v{\Omega}\v{k})^2\right]}
			 {\Omega^2 k^2 \left(k^2-k_r^2\right)}
		+ O\left(\frac{c_{\rm SFL}^4 k^4}{\mathcal{N}^4}\right)
.
\end{gather}
Here we see that the normal and superfluid modes, described by equations \eqref{eq:wkb-y-1} and \eqref{eq:wkb-y-2}
respectively, exhibit qualitatively different behaviour as $T \to T_{\rm cn}$:
$i^o$-mode frequencies vanish\footnote{
We remind the reader that we are considering only pure $i$-modes
with $\sigma \propto \Omega$ at low rotation frequencies;
in barotropic ($\mathcal{N}=0$) star normal inertial modes survive at $T > T_{\rm cn}$,
since they always have frequency $\sigma^2 = 4 (\v{\Omega}\v{k})^2/k^2$,
(see equations~\ref{eq:wkb-g-1} and \ref{eq:wkb-g-3} with $\mathcal{N}=0$).
In non-superfluid non-barotropic stars inertial modes (except for the single $r$-mode) do not exist,
since in the limit $\Omega \to 0$ they turn into g-modes \citep{Unno_etal89}. 
Therefore it is not surprising that the frequencies of $i^o$-modes tend to zero at $T \to T_{\rm cn}$.
},
whereas $i^s$-mode frequencies are finite.
One can clearly observe such behaviour in Fig.~\ref{fig:spectrumnpe} and Fig.~\ref{fig:spectrumnpe-lowTcp}.

To sum up, in this section we obtained dispersion relations for normal and superfluid inertial modes in superfluid $npe$ matter in a short wavelength limit.
We analyzed this relations in two opposite limiting cases.
The first case, $c_{\rm SFL} k / \mathcal{N} \gg 1$, describes behaviour of inertial modes at low temperatures $T \ll T_{\rm cn}$.
The corresponding relations are presented by equations~\eqref{eq:wkb-g-1}-\eqref{eq:wkb-g-2} if $h_1 \neq h$
and by equation~\eqref{eq:wkb-g-3} if $h_1 = h$.
These relations explain why the frequencies of superfluid $i^s$-modes are close to frequencies of the corresponding normal $i^o$-modes
if the entrainment effect is small or absent.
The second case, $c_{\rm SFL} k / \mathcal{N} \ll 1$, corresponds to the limit $T \to T_{\rm cn}$,
in which $i^o$-mode frequencies (equation~\ref{eq:wkb-y-1}) tend to zero, while $i^s$-mode frequencies (equation~\ref{eq:wkb-y-2}) remain finite.
These conclusions are consistent with the properties of the inertial modes spectrum calculated in Section~\ref{sec:numerical} (see Fig.~\ref{fig:spectrumnpe} and Fig.~\ref{fig:spectrumnpe-lowTcp}).

\section{Summary}
\label{sec:summary}

In this paper we studied the properties of inertial modes
in slowly rotating superfluid $npe$ NSs.
We calculated the spectrum of $l_0 - |m| = 1$, $l_0 - |m| = 3$, and $l_0 - |m| = 5$ inertial modes for $m=2$,
working in the leading order in rotation and including
both the entrainment and finite temperatures effects for the first time.
In Section~\ref{sec:numerical} we present the first results of such calculations.
We worked in the Cowling approximation, in the Newtonian limit (but employed relativistic EOS)
and assumed constant baryon critical temperatures throughout the core.

One of the motivations for 
doing this work was 
to find possible
avoided crossings in the plane `mode frequency -- stellar temperature'
between the normal $r^o$-mode and superfluid inertial modes ($i^s$-modes).
At stellar temperatures corresponding to  avoided crossings, eigenfunctions of $r^o$-mode and $i^s$-mode mix with each other. 
This stabilizes the $r^o$-mode 
and explains the existence of 
hot and rapidly rotating NSs in LMXBs \citep{gck14a,gck14b}.
In our simplified physical model we did not reveal any avoided crossing of $r^o$-mode with the superfluid modes in the leading order in the rotation frequency $\Omega$.
However, we found that at temperatures close to the critical neutron temperature 
one should expect stabilizing interaction of $r^o$-mode with the superfluid $r^s$-mode
(i.e., $i^s$-mode with $l_0 - |m| = 1$).
Moreover, we showed, for our particular stellar model, 
that at low enough temperatures $l_0 - |m| = 3$ $i^s$-mode has a frequency sufficiently close to that of $r^o$-mode and hence may also stabilize $r^o$-mode.

We also developed an approximate method which simplifies calculations  
of the $r^s$-mode eigenfrequency. 
This method is
similar to that used by \cite{kg17} and it allows one to calculate $r^s$-mode  analytically
in the limit of small entrainment (see Section~\ref{sec:r-mode}), assuming $npe$ core composition and working in the leading order in $\Omega$.
We found a good agreement of our analytical and numerical results. 
In addition, in Section~\ref{sec:wkb} we derived and discussed 
dispersion relations for normal and superfluid inertial modes 
in the short wavelength limit.
Using these relations we explained some properties of inertial modes
at low ($T \ll T_{\rm cn}$) and high ($T \to T_{\rm cn}$) stellar temperatures,
found numerically in
Section~\ref{sec:numerical}.

In the future we plan
to combine two analytical methods for calculating $r$-modes
(the method presented in Section~\ref{sec:r-mode} and the method by \citealt{kg17}),
accounting for both
entrainment effect
and next-to-leading order corrections in $\Omega$.
This will allow us to determine avoided crossings between the normal and superfluid $r$-modes
(and thus determine the $r$-mode instability windows) under more realistic assumptions.
We also plan to calculate the spectrum of inertial modes
for more elaborated neutron star models allowing for muons and 
density-dependent profiles of nucleon critical temperatures in the stellar core.

%
\section{Acknowledgments}
This work is partially supported by the Foundation 
for the Advancement of Theoretical Physics and Mathematics BASIS 
[Grant No. 17-12-204-1]
and by RFBR [Grant No. 18-32-20170].

\bibliography{literature}
\bibliographystyle{mnras}
\end{document}